\documentclass{raa}

\usepackage{graphicx,times}

\usepackage{natbib}

\bibpunct[, ]{(}{)}{;}{a}{}{,}

\def\beq{\begin{equation}}
\def\eeq{\end{equation}}
\def\bear{\begin{eqnarray}}
\def\ear{\end{eqnarray}}

\begin{document}
\title{X-ray Spectroscopy of Clusters of Galaxies}
 \volnopage{ {\bf 2012} Vol.\ {\bf 12} No. {\bf 8},~ 973--994}
   \setcounter{page}{1}

\author{Naomi Ota\inst{1}}
\institute{Department of Physics, Nara Women's University, Kitauoyanishimachi, Nara, Nara 630-8506, Japan; {\it naomi@cc.nara-wu.ac.jp}\\
\vs \no
   {\small Received --; accepted --}
}

\abstract{Clusters of galaxies are the most massive objects in the
  Universe and precise knowledge of their mass structure is important
  to understand the history of structure formation and constrain still
  unknown types of dark contents of the Universe. X-ray spectroscopy
  of galaxy clusters provides rich information about the
  physical state of hot intracluster gas and the underlying potential
  structure. In this paper, starting from the basic description of
  clusters under equilibrium conditions, we review properties of
  clusters revealed primarily through X-ray observations considering 
their thermal and dynamical evolutions. The future prospects of
  cluster studies using upcoming X-ray missions are also mentioned.
  \keywords{galaxies: clusters: general --- galaxies: intergalactic
    medium --- X-rays: galaxies: clusters --- Cosmology: observations
    --- Cosmology: dark matter}}

   \authorrunning{N. Ota}
   \titlerunning{X-ray Spectroscopy of Galaxy Clusters}
   \vspace{-3mm} \no{\sf INVITED REVIEWS}
   \maketitle

\section{Introduction}
According to the standard cosmological model, the Universe began 
13.8 billion years ago, and consists of 4\% baryonic matter, 23\% dark
matter (of unknown type) and 73\% dark energy (also of unknown origin)
\citep{larson11,komatsu11}.  Through interactions of these constituents, the
associated cosmic structures have been evolving up to now.  Our description of
the Universe is often based on the notion that large objects, like
galaxy clusters, that formed out of the evolving large-scale
structure, have attained an equilibrium state in their matter and
energy constituents.  However, is this truly a natural assumption?  To
tackle this problem, by focusing on objects appearing at the top of
the hierarchical structure formation, namely clusters of galaxies, is
vital in astrophysics.

\begin{figure}
    \centering
        \includegraphics[scale=0.49]{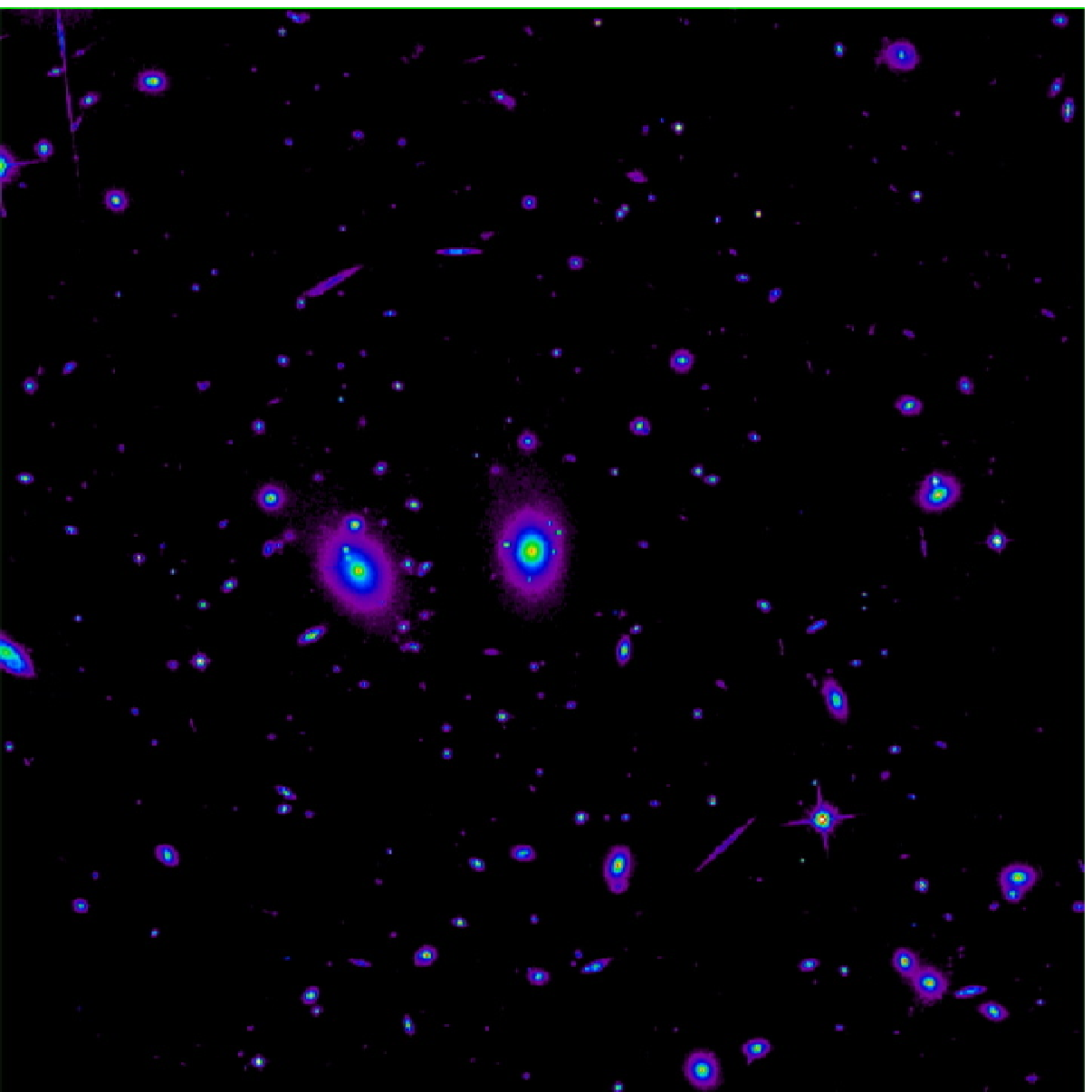}
        \includegraphics[scale=0.5]{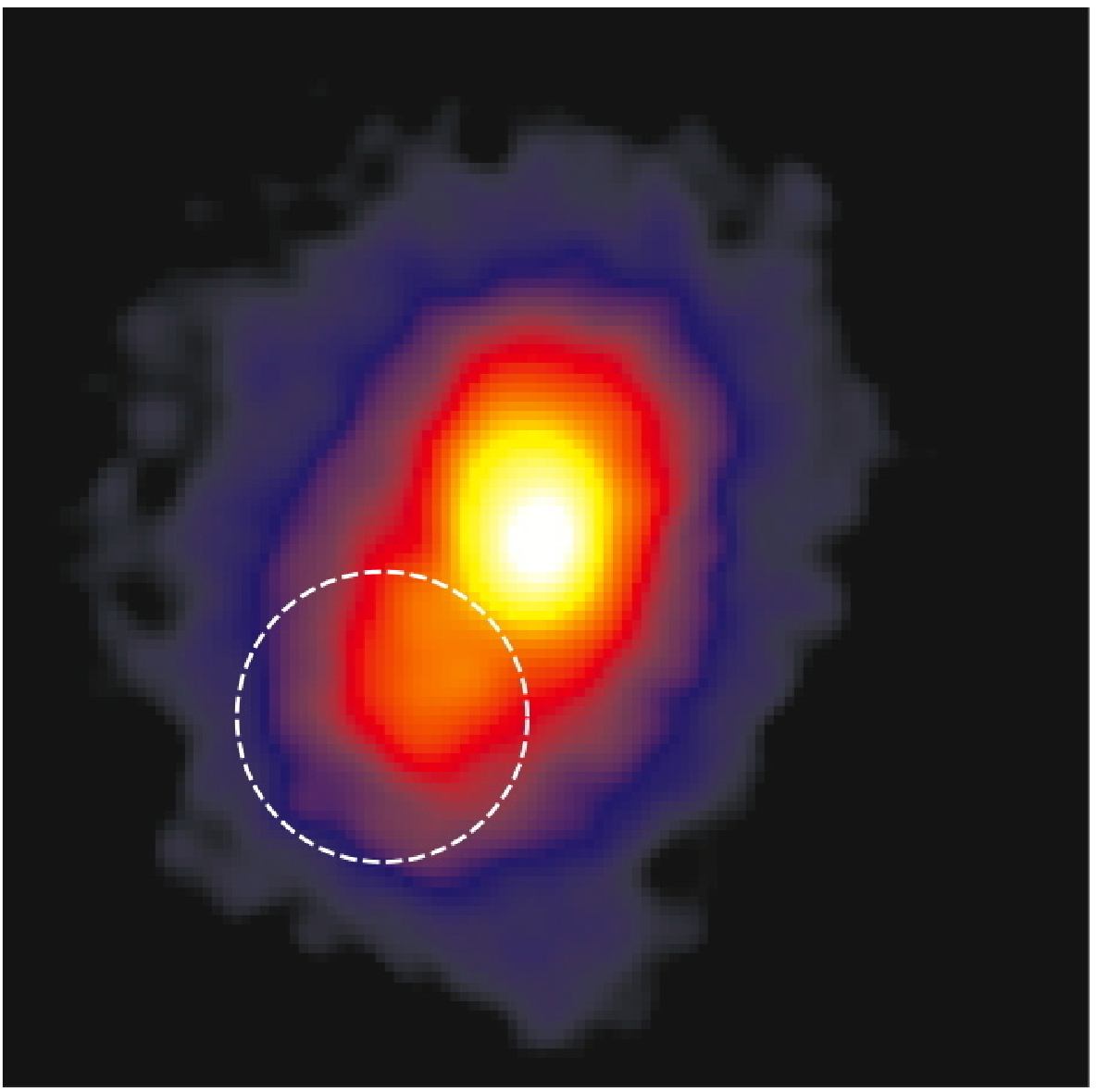}
        \caption{Optical (left) and X-ray (right) images of a cluster
          of galaxies, RX~J1347.5--1145, taken with the Hubble Space
          Telescope (the Multi-mission Archive at STScI) and the
          {\it Chandra} satellite, respectively. In both panels, a side of the figure is
          110\arcsec, corresponding to about 630~kpc. The white circle
          in the right panel indicates a location where extremely hot
          thermal gas has been discovered (see
          \S~\ref{subsec:super-hot}).  }
        \label{fig:rxj}
   \end{figure}

Clusters of galaxies are the largest gravitationally bound systems in
the Universe.  This makes them very important probes of cosmology.
Thus a precise knowledge of their mass structure is very important to
measure the large-scale structure and to test cosmological models.  In 
visible light, they are identified as groups of $\sim100-1000$
galaxies, extending over $\sim 10^7$ light years (Figure~\ref{fig:rxj}
left). On the other hand, X-ray observations of clusters have
drastically changed our view of cosmic structure: hot gas fills
inter-galactic space and emits strong X-rays (Figure~\ref{fig:rxj}
right). Furthermore, the total mass of hot gas exceeds the sum of
galaxy mass by two--three times.  To confine the hot gas by gravitational
forces, invisible matter, ``dark matter,'' of five times larger mass is
required.  As techniques in X-ray spectroscopy and imaging observations 
progressed, the presence of a complex temperature structure was also found
in the X-ray emitting gas. Those facts have revealed that clusters
preserve the past history of being built through complex
interactions, particularly merging, between smaller systems.  Thus the
clusters are no longer thought to be in an equilibrium state, but rather 
dynamically evolving on cosmological time scales.

This paper is organized as follows: in \S~2 a general description of
clusters is summarized. In \S~3--4, properties of clusters of galaxies
revealed primarily by X-ray observations are reviewed in light of
their thermal and dynamical evolutions.  Finally in \S 5, future
prospects are briefly mentioned. We use $\Omega_M=0.3$,
$\Omega_\Lambda=0.7$ and $h_{70}\equiv H_0/(70~{\rm
  km\,s^{-1}\,Mpc^{-1}})=1$ throughout the paper except where noted.

\section{Spatial distribution of matter and X-ray emission}
In the X-ray energy band, clusters of galaxies look very different
from the optical view; hot diffuse plasma with a temperature of $\sim
10^7-10^8$~K fills the intracluster space. The X-ray emitting hot
plasma is confined in the cluster gravitational potential and is
believed to trace the underlying dark matter distribution. The
typical X-ray luminosity of clusters is $10^{44\sim45}~{\rm
  erg\,s^{-1}}$, and the electron number density of hot plasma at
the center of clusters is typically $10^{-3\sim-2}~{\rm cm^{-3}}$.  In
what follows, the general view of clusters of galaxies under 
equilibrium models is summarized.

\subsection{Hydrostatic equilibrium condition and the $\beta$-model}
Since the collision time scales for ions and electrons in the
intracluster gas are much shorter than the time scales of heating or
cooling, we can treat the gas as a fluid \citep{sarazin88}. In general,
the sound crossing time $t_s$, i.e., the time required for a sound
wave in the intracluster gas to cross a cluster with radius $R$,
\begin{equation}
	t_s \equiv \frac{2R}{c_s} \sim 2~{\rm Gyr}\left(\frac{R}{1~{\rm Mpc}}\right) \left(\frac{c_s}{1000~{\rm km\,s^{-1}}}\right)^{-1}
\end{equation}
is shorter than the probable age of the cluster or the Hubble time,
$t_{H}=H_0^{-1}\sim14$~Gyr.  Thus the gas is considered to be in
hydrostatic equilibrium. In addition, if the cluster is spherically
distributed, the hydrostatic equation reads
\begin{equation}
\frac{1}{\rho_{\rm g}}\frac{dP_{\rm g}}{dr} = - \frac{d\phi}{dr} = -\frac{GM(r)}{r^2}, 
\end{equation}
where $M(r)$ is the total cluster mass (i.e., dark matter +
galaxies + hot gas) within the radius $r$ and $P_{\rm g}$ is the
thermal pressure and a product of gas density and temperature, $n_{\rm
  g}(r) kT_{\rm g}(r)$. If the self-gravity of the gas is ignored, the
distribution of gas is determined by the cluster potential,
$\phi$.

A temperature gradient in the plasma is smoothed by heat
conduction. If the heat conduction were sufficiently rapid compared to
other important time scales, the gas would become
isothermal. Substituting the gas pressure $P(r)=n_{\rm g}(r) T_{\rm
  g}$ into Eq. (1) and assuming $T_{\rm g}$ is constant, we obtain
\begin{equation}
	\frac{d\ln{\rho_{\rm g}}}{dr} = - \frac{\mu m_p}{kT_{\rm g}}\frac{d\phi}{dr}, 
\end{equation}
where $\mu$ is the mean molecular weight, $\sim 0.6$. Similarly,
the galaxies are bounded in the gravitational potential, whose
hydrostatic condition is written as
\begin{equation}
	\frac{d\ln{\rho_{\rm G}}}{dr} = - \frac{1}{\sigma^2}\frac{d\phi}{dr}.
\end{equation}
$\sigma$ is the line-of-sight velocity dispersion and is typically of
the order of 1000~${\rm km\,s^{-1}}$. From Eqs. (2) and (3), we find
\begin{equation}
	\rho_{\rm g} = \rho_{\rm G}^{\beta}, ~~~\beta\equiv\frac{\mu m_p \sigma^2}{kT_{\rm g}}.
\end{equation}
Hence the gas distribution differs just by the index $\beta$ in
comparison with that of member galaxies.

\citet{king62} derived an analytic approximation to the isothermal
sphere of self-gravitational isothermal collision-less particles. The
density profile of the cluster's member galaxies has been found to be
well approximated by the King profile,
\begin{equation}
	\rho_{\rm G}\sim \rho_{\rm King} = \rho_0 \left[ 1 + \left(\frac{r}{r_c}\right)^2 \right]^{-3/2}.
\end{equation}
Here $r_c$ represents a core radius within which the density is regarded
as constant. From Eqs. (4) and (5), the isothermal gas distribution is represented by: 
\begin{equation}
	\rho_{\rm g} = \rho_{\rm g0} \left[ 1 + \left(\frac{r}{r_c}\right)^2 \right]^{-3\beta/2}.\label{eq:density_gas}
\end{equation}
This formula is called the isothermal $\beta$-model \citep{cavaliere76}.

\subsection{X-ray emission process}
An X-ray spectrum emitted from an ionized intracluster plasma is
described with a combination of continuum emission and line emission
from heavy elements.  The former is produced by free-free (or
bremssstrahlung), free-bound, two-photon emission and the latter is by
one-electron radiative transitions, dielectric recombination satellite
lines, and inner-shell ionization \citep{vanparadijs99,bohringer10}.

In the temperature range of clusters ($kT>2$~keV), the total emission
is dominated by the free-free emission if the abundance of heavy
elements does not exceed the solar abundance by very much. The
emissivity of the free-free emission at a frequency $\nu$ from a hot
plasma with an electron temperature of $T_{\rm g}$ is given by
\begin{eqnarray}
\epsilon^{\rm ff}_{\nu} &=& \frac{2^5\pi e^6}{3m_ec^3}(\frac{2\pi}{2m_ek})^{1/2} 
n_e \sum Z^2n_i g_{\rm ff}(Z,T_g,\nu) \times T^{-1/2}_g\exp{(-h\nu/kT_g)} \label{eq:bremss}\\
 &=& \Lambda(T,Z,\nu) n_e^2
\end{eqnarray}
where $Z$ is an ion of charge in a plasma, $n_i$ and $n_e$ are the
number density of ions and electrons, respectively
\citep[e.g.,][]{rybicki86}.  The Gaunt factor is a correction factor
for quantum mechanical effects and is approximately $g_{\rm ff}\sim
0.9(h\nu/kT)^{-0.3}$.  The bolometric emissivity is then
\begin{eqnarray}
\epsilon^{\rm ff} &=& \int^{\infty}_{0} \epsilon^{\rm ff}_{\nu} d\nu  =  \Lambda(T,Z) n_e^2 \nonumber \\ 
 &\sim& 1.435\times10^{-27} \bar{g} T_g^{1/2} n_e\sum Z^2n_i ~[{\rm erg\,s^{-1}\,cm^{-3}}]. \label{eq:bremss_emissivity}
\end{eqnarray}

\begin{figure}
    \centering
        \includegraphics[scale=0.4]{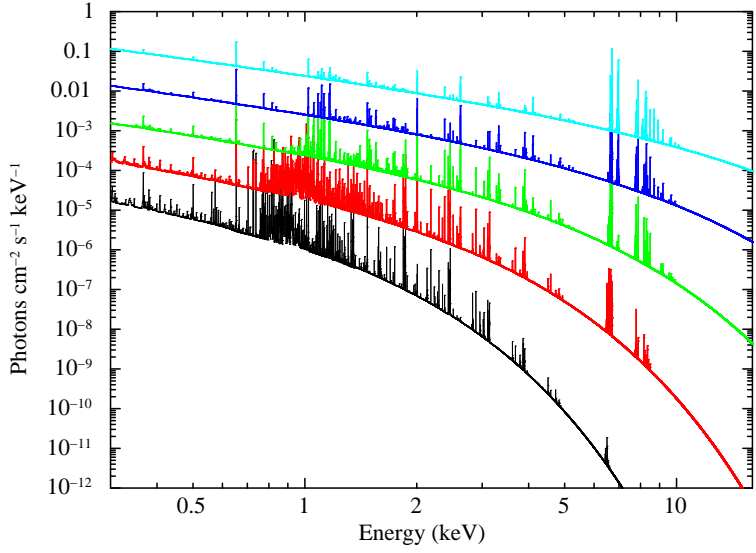}
        \caption{APEC thin-thermal plasma model for $kT/{\rm
            [keV]}=$1(black), 2(red), 4(green), 8(blue), 16(cyan), and
          the metal abundance of 0.3~solar. The graphs are shifted in
          the $y$-direction for clarity.}
        \label{fig:apec}
   \end{figure}

Precise X-ray emission spectra from thin-thermal plasma can be
calculated by utilizing plasma codes such as APEC \citep{smith01}, and 
MEKAL \citep{mewe85, mewe86,kaastra92,liedahl95}.  The updated version
of the latter is available in the SPEX package \citep{kaastra96}.  For
reference, the APEC models for various temperatures are plotted
in Fig.~\ref{fig:apec}. The metal abundance is assumed to be
0.3~solar, as is typical of intracluster gas \citep{mushotzky97}. The
abundance table of \citet{anders89} is used here\footnote{The updated
  table for the solar system abundance is given in \citet{lodders03}}.

\subsection{$\beta$-profile and hydrostatic mass estimate} 
The surface brightness profile of an isothermal spherical plasma with
a radial density profile given by Eq.\ref{eq:density_gas} is
calculated by integrating the local emission per unit volume
(Eq.\ref{eq:bremss_emissivity}) and the density along the line of
sight.  We obtain the X-ray surface brightness $S(r)$ at a projected
radius $r$,
\begin{eqnarray}
S(r) &=& S_0 \left[1+\left(\frac{r}{r_c}\right)^2\right]^{-3\beta+1/2}, \label{eq:beta_sb}\\
S_0 &\equiv& n_{e0}n_{\rm H0} \Lambda(T,Z)
\frac{\sqrt{\pi} r_c}{4\pi D_L^2}\frac{\Gamma(3\beta-1/2)}{\Gamma(3\beta)} 
\,{\rm [erg\,s^{-1}\,cm^{-4}]}\label{eq:beta_sb_center}. 
\end{eqnarray}
Here $n_{e0}$ and $n_{H0}$ are the central electron 
and hydrogen densities of the intracluster gas respectively and $D_L$ is a luminosity distance to the object.
It is known that the observed cluster's X-ray surface brightness is well
fitted with the above function, and $\beta \sim 0.6-0.7$ on average
\citep[e.g.,][]{jones84,ota04b}.

Once we have obtained the $\beta$ profile parameters to characterize
the surface brightness distribution, we can estimate the
three-dimensional density profile of the gas, i.e. the $\beta$-model.
Then the mass of the gas inside a radius $r$ is given by integrating
Eq.\ref{eq:density_gas}.
\begin{equation}
M_{\rm gas}(r) 
 = 4\pi \rho_{\rm gas}(0) {r_c}^3 \int^{x}_0 (1+x^2)^{-3\beta/2}x^2 dx,
 \label{eq:beta_gasmass}
\end{equation}
where $x = r/r_c$. From the assumption of hydrostatic equilibrium, we
derive the total mass of the cluster inside a radius $r$,
\begin{equation}
M(r) = \frac{3kT\beta r}{\mu m_p G}\frac{(r/r_c)^2}{1+(r/r_c)^2} .
\label{eq:beta_totalmass}
\end{equation} 
If gas is not isothermal and its temperature distribution has a radial
dependence, $T_{\rm g}(r)$, the hydrostatic mass is rewritten as
\begin{equation}
	M(r) = - \frac{kT_{\rm g}(r)r}{G\mu m_p}\left[ \frac{\partial \ln{n_{\rm g}(r)}}{d\ln{r}} + \frac{\partial\ln{T_{\rm g}(r)} }{\partial \ln{r}}\right].
\end{equation}

\begin{figure}
    \centering
        \includegraphics[scale=0.25]{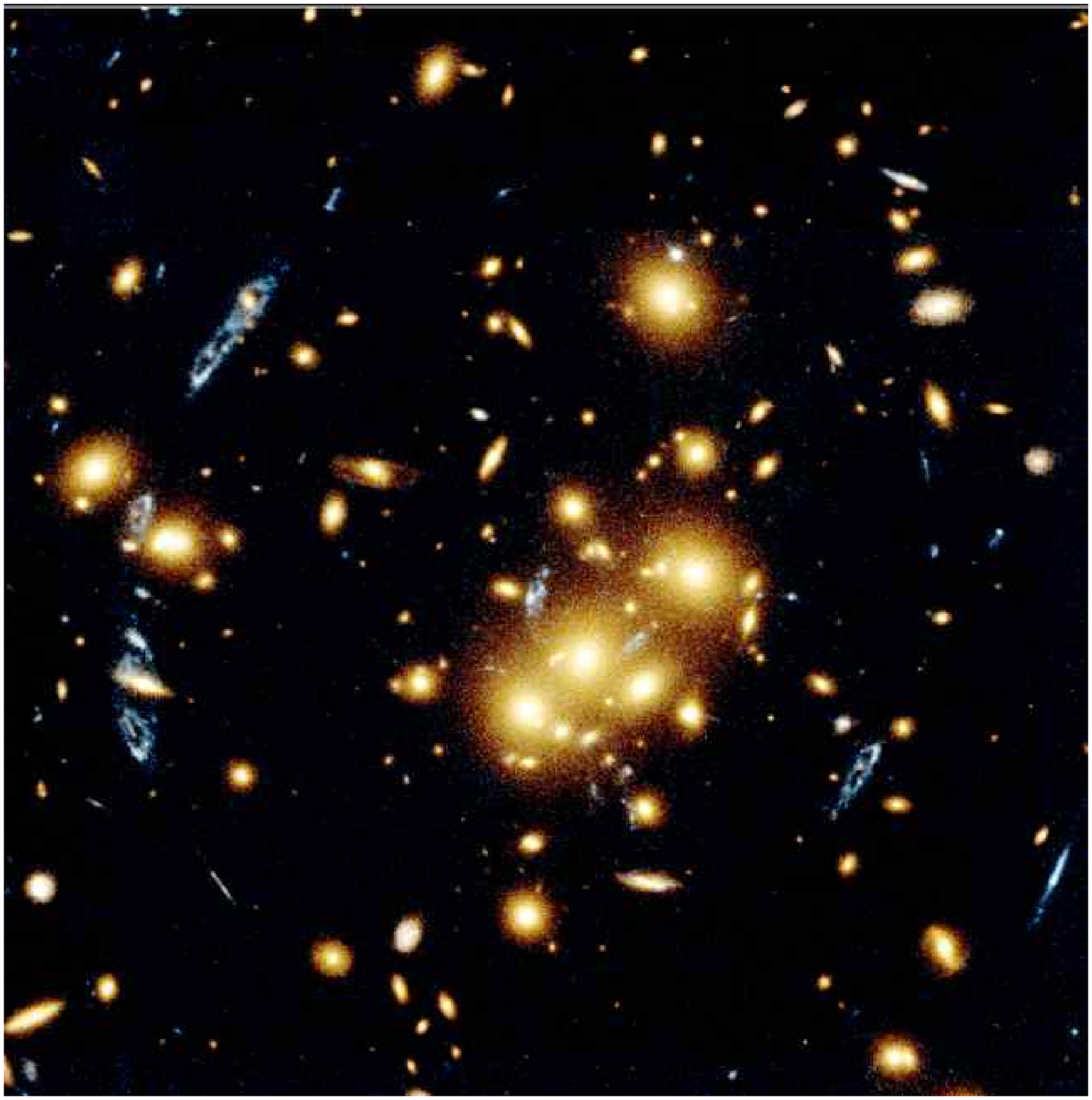}
        \includegraphics[scale=0.35]{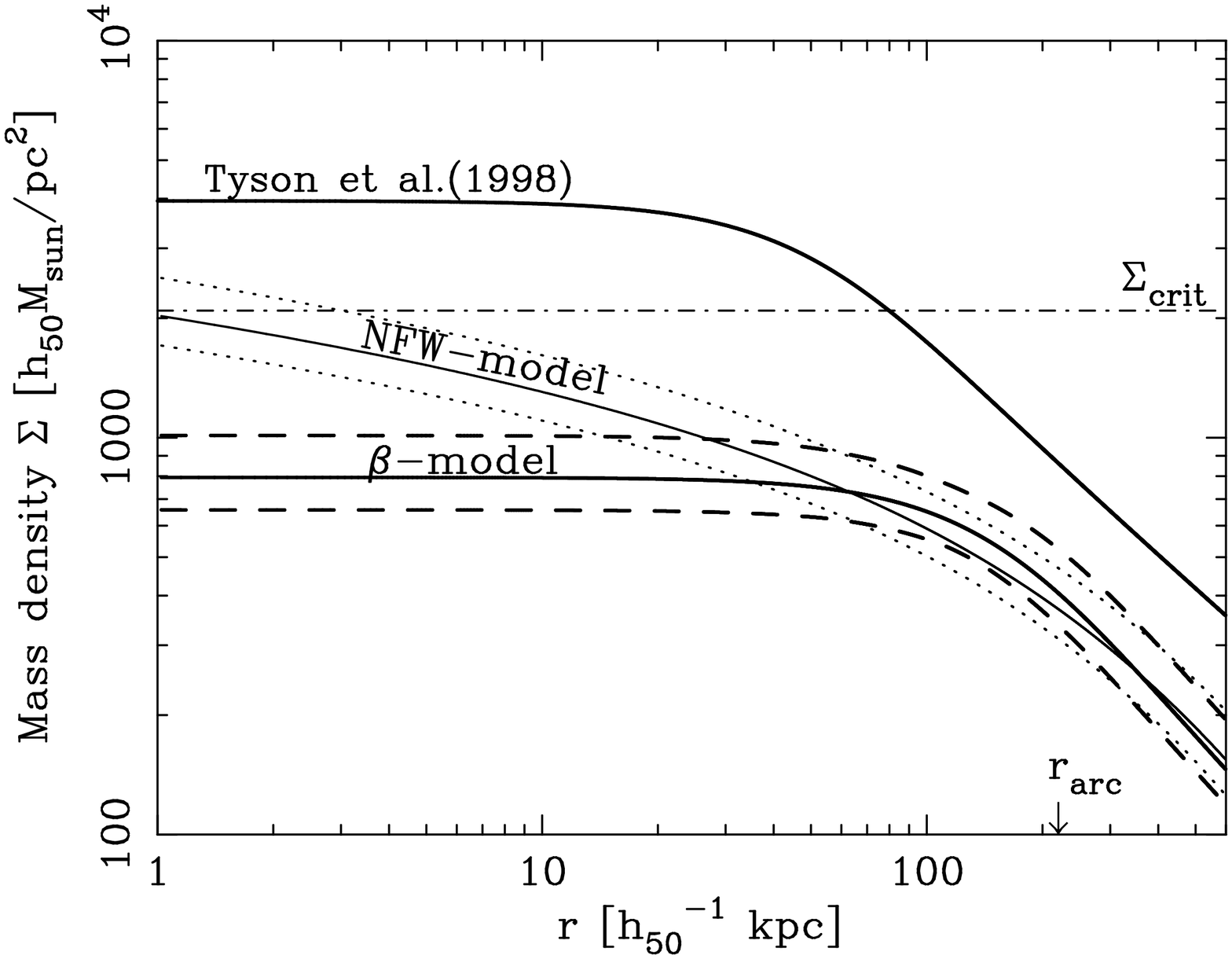}
        \caption{HST/WFPC2 image of a lensing cluster with multiple
          lens arcs, CL0024+17 (left). Mass density profiles for the
          $\beta$-model and the NFW-model of the CL0024+17 main
          cluster \citep{ota04a} are shown and compared with the
          lensing mass model by \citet{tyson98} (right).}
        \label{fig:cl0024}
   \end{figure}

As an illustration, the result of hydrostatic mass estimation under
the isothermal $\beta$-model for a gravitational lensing cluster
CL0024+17 ($z=0.395$) is shown in Fig.~\ref{fig:cl0024} and compared
with an independent mass determination based on the gravitational
lensing effect \citep{tyson98}. Since the lensing effect directly maps
the surface mass density of the cluster, regardless of the internal
dynamical and thermal state of the cluster, comparison of the two
methods provides information on the physical state of clusters
\citep[][for review]{hattori99,kneib11}. In the case of CL0024+17, a
factor of two--three discrepancy has been found between the hydrostatic and
strong lensing mass estimates, indicating that the system is
experiencing a line-of-sight merger
\citep{ota04a,zhang05,jee07,zuhone09}.

\subsection{Universal dark matter density profile}
\citet{navarro97} found from their numerical simulations of structure
formations under the Cold Dark Matter (CDM) model, that the collapsed
dark matter halos with masses over several orders of magnitude follow
a universal density profile,
\begin{equation}
\rho_{\rm NFW}(r)=\frac{\rho_s}{(r/r_s)(1+r/r_s)^2} , \label{eq:univ_prof}
\end{equation} 
where $\rho_s$ and $r_s$ are the characteristic density and length,
respectively.  $\rho_s$ is related to the critical density of the
universe $\rho_{\rm crit}$ and the characteristic density $\delta_c$
through $\rho_s = \delta_c \rho_{\rm crit}$. Instead of the flat core
of the King profile, the NFW profile has a core with $\propto r^{-1}$
dependence. Although the density diverges at the center, the mass
inside a radius $r$,
\begin{equation} 
M_{\rm NFW}(r) = 4\pi \rho_s{r_s}^3 \left[\ln(1+x) - \frac{x}{1+x}\right], 
\, x\equiv r/r_s\label{eq:mass_nfw}
\end{equation} 
converges to 0 as $r \rightarrow 0$.

The density distribution of intracluster gas in hydrostatic
equilibrium with the NFW dark matter potential was analytically derived
by \citet{makino98} assuming the masses of gas and galaxies are
negligibly small compared to the dark matter.
\begin{eqnarray}
\rho_{\rm gas}(r)&=&\rho_{\rm gas0}\exp\left[-B\left(1-\frac{\ln{(1+x)}}{x}\right)\right],\,\,\,\,B \equiv \frac{4\pi G \mu m_p \rho_s r_s^2}{kT} \label{eq:nfw_gas_prof}
\label{eq:makino}
\end{eqnarray} 
Then the mass of the gas within a radius $r$ is given by,
\begin{equation}
M_{\rm gas}(r) = 4\pi \rho_{\rm gas}(0) e^{-B}r_s^3\int_0^{x} x^2(1+x)^{B/x} 
dx. \label{eq:mgas_nfw}
\end{equation}

\citet{suto98} generalized the universal density profile to the form
$\rho\propto 1/[x^{\mu}(1+x^{\nu})]^{\lambda}$ and numerically
computed the gas density profile in hydrostatic equilibrium for the
case of $\mu=\alpha$, $\nu=1$, and $\lambda=3-\alpha$ with the
restriction $1<\alpha<2$. Note that the case with $\alpha=1$, $\nu=1$,
and $\lambda=2$ corresponds to the NFW model.  They further computed
the X-ray surface brightness distribution at a projected radius $r$ on
the sky, and derived a useful fitting formula in the following
generalized shape.
\begin{eqnarray} 
S(r) &\propto& \left[1+ \left(\frac{r}{r_s\phi_c}\right)^{\zeta}\right]^{-\eta} \\
\phi_c &=& 0.3(2/\alpha-1) \nonumber \\
\zeta  &=& 0.41-5.4(2-\alpha)^6+(0.585+6.47\alpha^{-5.1B})B^{-\alpha^
6/30} \nonumber \\
\eta    &=& -0.68-5.09(\alpha-1)^2+(0.202+0.0206\alpha^8)B^{1.1} \nonumber
\label{eq:nfw-suto}
\end{eqnarray}
These are valid for $5\leq B\leq20$ and $1.0\leq\alpha\leq1.6$ in the
range $10^{-4}\leq \phi \leq \phi_{\rm max}$, where $S(\phi_{\rm
  max})=10^ {-4}S(0)$.  We refer to the formula with $\alpha=1$ as the
SSM model hereafter.

The SSM model has a surface brightness distribution similar to the
$\beta$ profile over a wide range of $r$, although it has an excess
over the $\beta$-profile in the central region because of the strong
concentration of the dark matter halo of the NFW model.
\citet{makino98} fitted simulated gas profiles which obey the
universal dark matter profile with the $\beta$-profile function, and
noted that the best-fit relation between the scale parameter and the
$\beta$-model core radius is given by $r_c = 0.22 r_s$.

Two kinds of density profiles, the $\beta$-model and the SSM model, have been 
introduced so far since they give reasonable approximations to
observed gas profiles in studying the global cluster structure.
Deviation from those models sometimes seen at the center of clusters
will be mentioned later.

\subsection{Formation of clusters and the virial radius}
Numerical simulations based on the CDM model predict hierarchical
structure formation, so rather continuous accretion of matter and
sub-cluster merging occur in the process of cluster formation
\citep[e.g.,][]{moore01}. Hence clusters reside at junctions of cosmic
filaments and are connected to the surrounding filamentary structures.  It
is, however, practically important to define a `cluster' based on some
simple model.  In this section we briefly review the collapse scenario
according to the spherical collapse model \citep{gunn72}. This model
predicts a very important physical quantity of clusters, the virial
radius.

At some time epoch, a certain region of the Universe which happens to
have a higher mass density than the background due to fluctuations
starts breaking away from the general expansion, and eventually
collapses to form a cluster of galaxies.  Since at the collapse epoch,
$\Omega \sim 1$, we can neglect the $\Lambda$ term in the equation of
motion of the shell.  We also assume that the amplitude of the density
perturbation is small, i.e $\delta \ll1$.  Then we have
\begin{equation}
\frac{d^2 r}{d t^2} = -\frac{{\rm G}M}{r^2}
\end{equation}
where $M$ is the mass inside the shell and is constant.
The first integral of this equation is,
\begin{equation}
\left(\frac{dr}{dt}\right)^2 = \frac{2{\rm G}M}{r} + C .
\label{eq:collapse_1st_integral}
\end{equation}
$C$ is a constant, and the total energy $C/2$ must be negative for
collapse to occur.  The solution of
Eq.~(\ref{eq:collapse_1st_integral}) is given in a parametric form,
\begin{equation}
t = \frac{{\rm G}M}{|C|^{3/2}}(\theta - \sin \theta), \,\,\, r = \frac{{\rm G}M}{|C|}(1 - \cos \theta) .
\end{equation}
The radius, $r$, is 0 at $\theta=0$, i.e. $t=0$.  Then it increases
with increasing $\theta$ and takes the maximum,$ r_{\rm m} = 2{\rm
  G}M/|C|$ at $\theta=\pi$, i.e. , $t=t_{\rm m} \equiv \pi {\rm
  G}M/|C|^{3/2}$ (turn around). Then it shrinks to 0 again at
$\theta=2\pi$, i.e., $t=t_{\rm c} \equiv 2\pi {\rm G}M/|C|^{3/2}$
(collapse).  After collapse, the system will be virialized.  In the
virialized system, the potential energy is related to the total energy
as $W = 2E$. Assuming the radius of the system after virialization is
$r_ {\rm vir}$, we have
\begin{equation}
W = - \frac{{\rm G}M^2}{r_{\rm vir}} = 2E = -2 \frac{{\rm G}M^2}{r_{\rm m}}. 
\end{equation}
Therefore, $r_{\rm vir} = r_{\rm m}/2$.  The average density inside
the virial radius $r_{\rm vir}$ is
\begin{equation}
\bar{\rho}_{\rm vir} = \frac{3}{4\pi} \frac{|C|^3}{{\rm G}^3 M^2}.
\end{equation}

On the other hand, the solution of
Eq.~(\ref{eq:collapse_1st_integral}) with $C=0$ describes the
background expansion, because $\Omega \sim 1$.  The solution is
\begin{equation}
r_{\rm b} = \left(\frac{9}{2}{\rm G}M\right)^{1/3} t^{2/3} .
\end{equation}
The density inside $r_{\rm b}$ gives the critical density at $t$,
\begin{equation}
\rho_{\rm crit}(t=t_{\rm c}) = \frac{1}{24\pi^3} \frac{|C|^3}{G^3 M^2}.
\end{equation}
Thus we obtain the important relation,
\begin{equation}
\Delta_c \equiv \frac{\bar{\rho}_{\rm vir}}{\rho_{\rm crit}} = 18 \pi^2. \label{eq:Deltac_ES}
\label{eq:rvir}
\end{equation}
We can assume that a cluster is virialized within the overdensity
radius $r_{\Delta}$ at which the average density is equal to
$\Delta_c$ times the critical density of the collapsed
epoch. 

The spherical collapse in an $\Omega + \Lambda = 1$ Universe is presented
in the Appendix of \citet{nakamura97}: a fitting formula for the
overdensity in the flat Universe with finite $\Lambda$ is
\begin{equation}
\Delta_c \simeq 18\pi^2\Omega^{0.437}.
\end{equation}

By taking $\Delta_c = 180$ or 200, the overdensity radius of $r_{180}$
or $r_{200}$ is often quoted as a measure of the cluster's virial radius.
$r_{500}$ is also frequently used for the reason that there is an
indication from numerical simulation that the hydrostatic assumption
is valid within that radius \citep{evrard96} as well as that X-ray
signals being detected out to $r_{500}$ or roughly $\sim 1$~Mpc in many
clusters (beyond that deeper exposure is required to trace emission
from the tenuous matter). The temperature scaling for the overdensity
radii for various $\Delta$ is derived using a nearby X-ray cluster
sample \citep{arnaud05}: for $\Delta_c=500$, it resulted in
$r_{500}h(z) = (1104 \pm 13)(kT/5~{\rm keV})^{0.57\pm 0.02}$~kpc.

\subsection{Radiative cooling of gas}\label{subsec:tcool}
Since hot intracluster gas loses its thermal energy via X-ray
emission, radiative cooling may affect the cluster structure once the
gas is settled in the cluster's potential.

The thermal energy loss is expressed by
\begin{equation}
\frac{d E_e}{dt} = - \epsilon^{\rm ff},
\end{equation}
where $E_e$ is the thermal energy of electrons per unit volume and $E_e
= 3 n_e kT_g/2$.  The volume emissivity, $\epsilon^{\rm ff}$ can be
denoted as $\epsilon^{\rm ff} = q_{\rm ff} n_e^2 T_g^{1/2}$.  Thus if
the hot gas cools, keeping the density constant, the temperature
decreases according to the following equation
\begin{equation}
\frac{d T_g}{dt} = -a T_g^{1/2}, ~~~ a = \frac{2 q_{\rm ff} n_e}{3 k}.
\end{equation}
The solution is
\begin{equation}
T_g(t)^{1/2} = T_g(0)^{1/2} -  \frac{a}{2} t .
\end{equation}
Thus the hot gas cools on the time scale
\begin{eqnarray}
t_{\rm cool} & =& \frac{2 T_g(0)^{1/2}}{a}= \frac{3k T_g(0)^{1/2}}{q_{\rm ff} n_{e}} \\ \label{eq:tcool}
	& \sim & 3\times10^{9}~{\rm yr}\left(\frac{T_g(0)}{4~{\rm keV}}\right)^{1/2}\left(\frac{n_e}{2\times10^{-2}~{\rm cm^{-3}}}\right)^{-1} 
\end{eqnarray}
The cooling time scale for the central region of typical relaxed
clusters is estimated to be shorter than the Hubble time, the cluster
core may be subject to radiative cooling. On the other hand, the
radiative cooling is not considered to be important outside the core
region because of lower density.

\section{Thermal evolution of intracluster gas}
\subsection{Cooling problem}
According to \S~\ref{subsec:tcool}, gas at the cluster's center can
radiate an amount of energy comparable to its total thermal energy
in less than the Hubble time and thus cools. It was suggested from
earlier works that the ``cooling-flow'' phenomenon would occur if the
gas cools isobarically and no heating process balances this cooling,
so the gas flows inward maintaining the thermal pressure \citep{fabian94}. 
On the other hand, high-resolution X-ray
spectroscopy showed that the temperature drops in the cooling cores by only
a factor of two--three, and there is much less emission at low
temperature, as predicted by the standard cooling-flow model
\citep[e.g.,][]{kaastra04}. This observational finding triggered explorations into a variety of scenarios for gas heating: heat conduction, 
active galactic nucleus (AGN) heating, magnetic reconnections, cosmic-ray heating etc.
X-ray and radio observations have
provided evidence for the interaction of AGN jets with cluster gas
\citep[e.g.,][]{mcnamara00}.  Although the work done by uplifting AGN
bubbles on the surrounding gas may be of the order of magnitude to
compensate the radiation loss, how the feedback achieves a tuning
between cooling and heating is not clear. The similarity and
smoothness of cooling profiles indicate the need for a continuous,
distributed heat source \citep[for review, e.g., ][]{peterson06}.

\subsection{Statistical properties of cluster cores}

\begin{figure}
    \centering
        \includegraphics[scale=0.35]{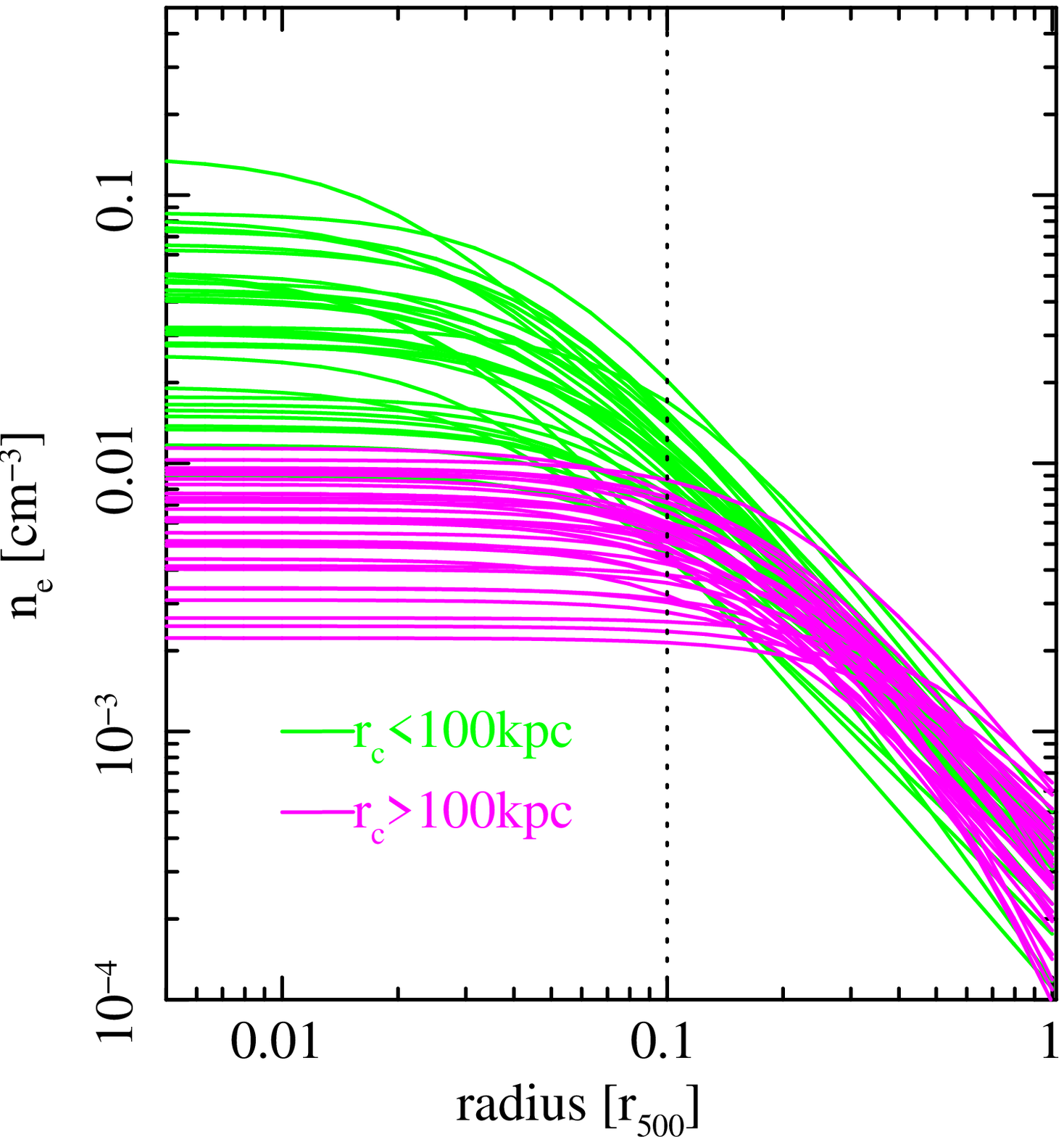}
        \includegraphics[scale=0.38]{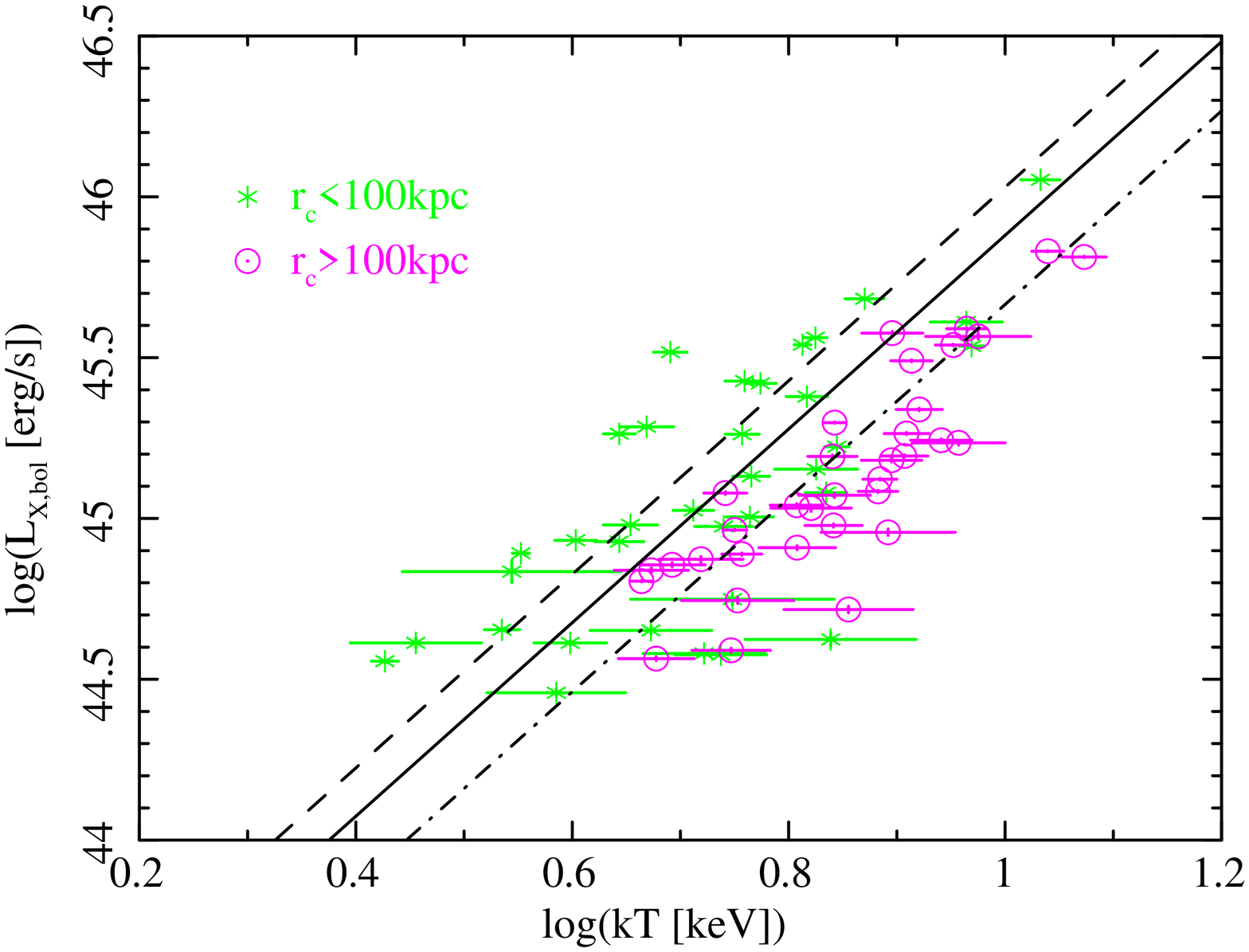}
        \caption{(Left) Electron density profiles for 69 clusters. The
          best-fit density profile derived with the single
          $\beta$-model are plotted, where the radius is normalized
          with $r_{500}$. $0.1r_{500}$ is indicated with the vertical
          dotted line, inside which the scatter is the most
          prominent. (Right) $L_{X}-T$ relation of
          clusters. A significant offset in the normalization factor of
          the $L_{X}-T$ relation between clusters with small
          ($r_c<100$~kpc) and large core radii ($>100$~kpc) is seen \citep{ota06}.}
        \label{fig:density}
\end{figure}

Regarding the density profile, a deviation from the conventional
isothermal $\beta$-model is commonly seen at the center of clusters
having a compact core (often termed Cool Core (CC) clusters): they
exhibit systematically higher central density while the profiles are
fairly universal outside $0.1r_{500}\sim100$~kpc \citep{neumann99}.
Fig.~\ref{fig:density} shows the gas density profiles derived with the
single-$\beta$ model \citep{ota06}.
The density scatter is prominent within $\sim 0.1r_{500}$ and is found
to be a significant source of scatter in the X-ray
luminosity-temperature correlation \citep{ota06,ohara06,chen07}.

\begin{figure}
    \centering
        \includegraphics[scale=0.4]{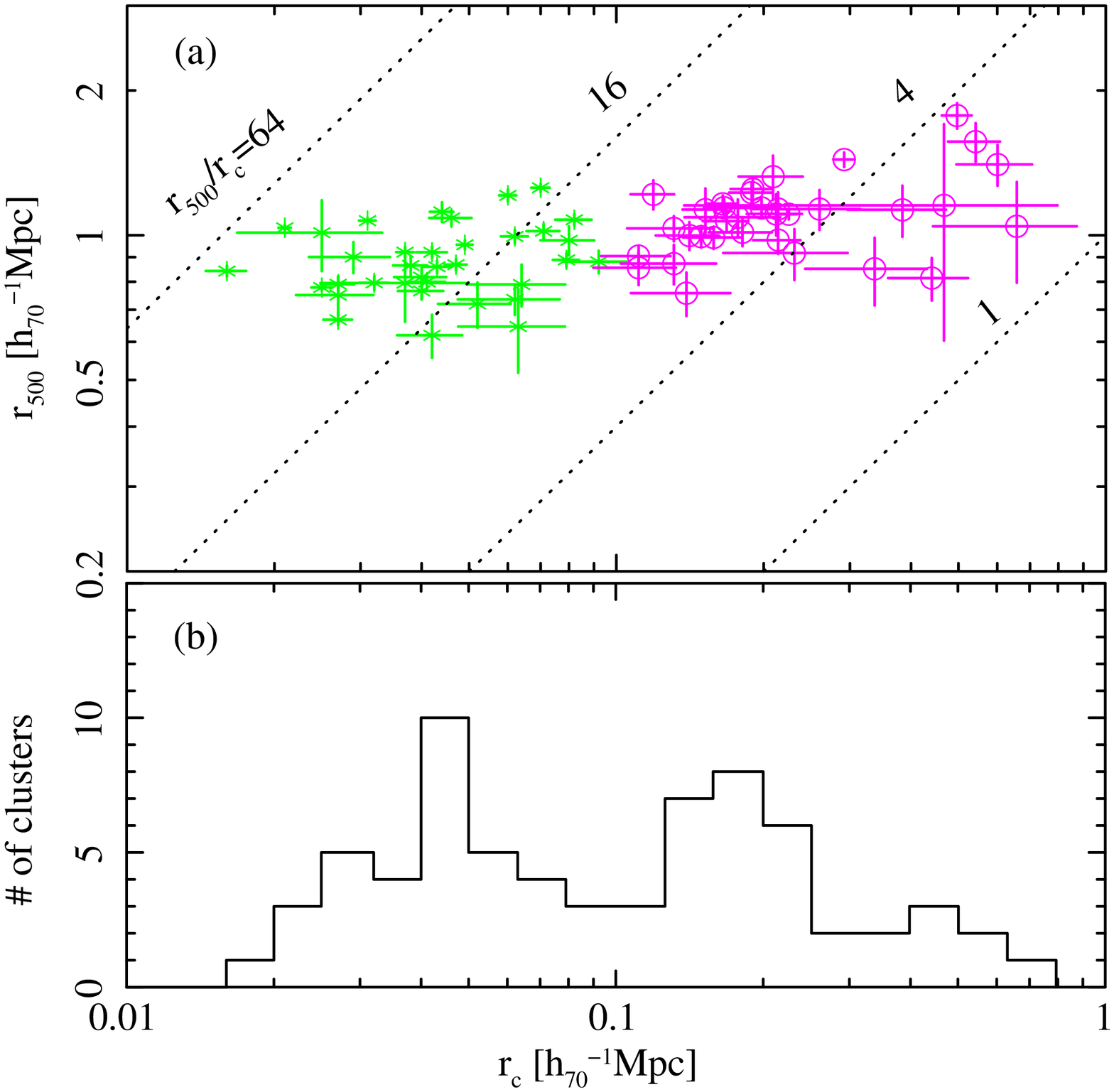}
        \caption{(a) $r_{500} - r_c$ relation and (b) histogram of
          $r_c$ for 69 clusters at $z>0.1$. In panel (a), 35
          clusters with a small core of $r_c <100$~kpc and 34 clusters
          with a larger core of $r_c>100$~kpc are shown with the
          asterisks and the circles, respectively. The dotted lines
          indicate the self-similar condition corresponding to four
          different constant values of $r_{500}/r_c$. }
        \label{fig:rc_hist}
\end{figure}

The statistical properties of gas density structure have been
investigated from systematic analysis of cluster samples by many
authors \citep[e.g.,][]{ohara06,ota06,chen07,santos08,cavagnolo09,
  hudson10}. These X-ray studies show that the fraction of CC clusters
is roughly 50\%. The rest of the sample without the central cool
emission is called Non-Cool Core (NCC) clusters.  \citet{ota02,ota04b}
first pointed out from the analysis of {\it ROSAT} and {\it ASCA}
archival data that the histogram of the cluster core radius exhibits a 
high concentration around 50 and 200~$h_{70}^{-1}$kpc
(Fig.~\ref{fig:rc_hist}). Later, a similar double-peaked distribution
of core radius was shown independently by \cite{hudson10}: they
utilized the {\it Chandra} data set on a nearby flux-limited sample
with higher resolution. The consistency between the two results gives
a confirmation of this nature.

The relaxed clusters often host a central dominant elliptical galaxy, also called a 
cD galaxy, which deepens the cluster potential well and causes a
peaked gas profile \citep{ikebe99}.  The regular clusters with a small
core tend to contain a cD galaxy, however, not all of them have one. Thus it is unlikely that the small core represents the
potential distribution of the cD galaxy itself \citep{akahori05}.

Under the self-similar model, the internal structure of the gas should
be scaled by the virial radius, and then $r_{500}/r_c$ should be
constant for all clusters. However, $r_c$ does not simply scale by
$r_{500}$ (Fig.~\ref{fig:rc_hist}), particularly for those having a 
small core radius ($r_c<100$~kpc). This clear departure from the
self-similar relation for small-core clusters suggests
that the formation of the small cores is determined by some physical
process other than the self-similar collapse.

\begin{figure}
    \centering
        \includegraphics[scale=0.3]{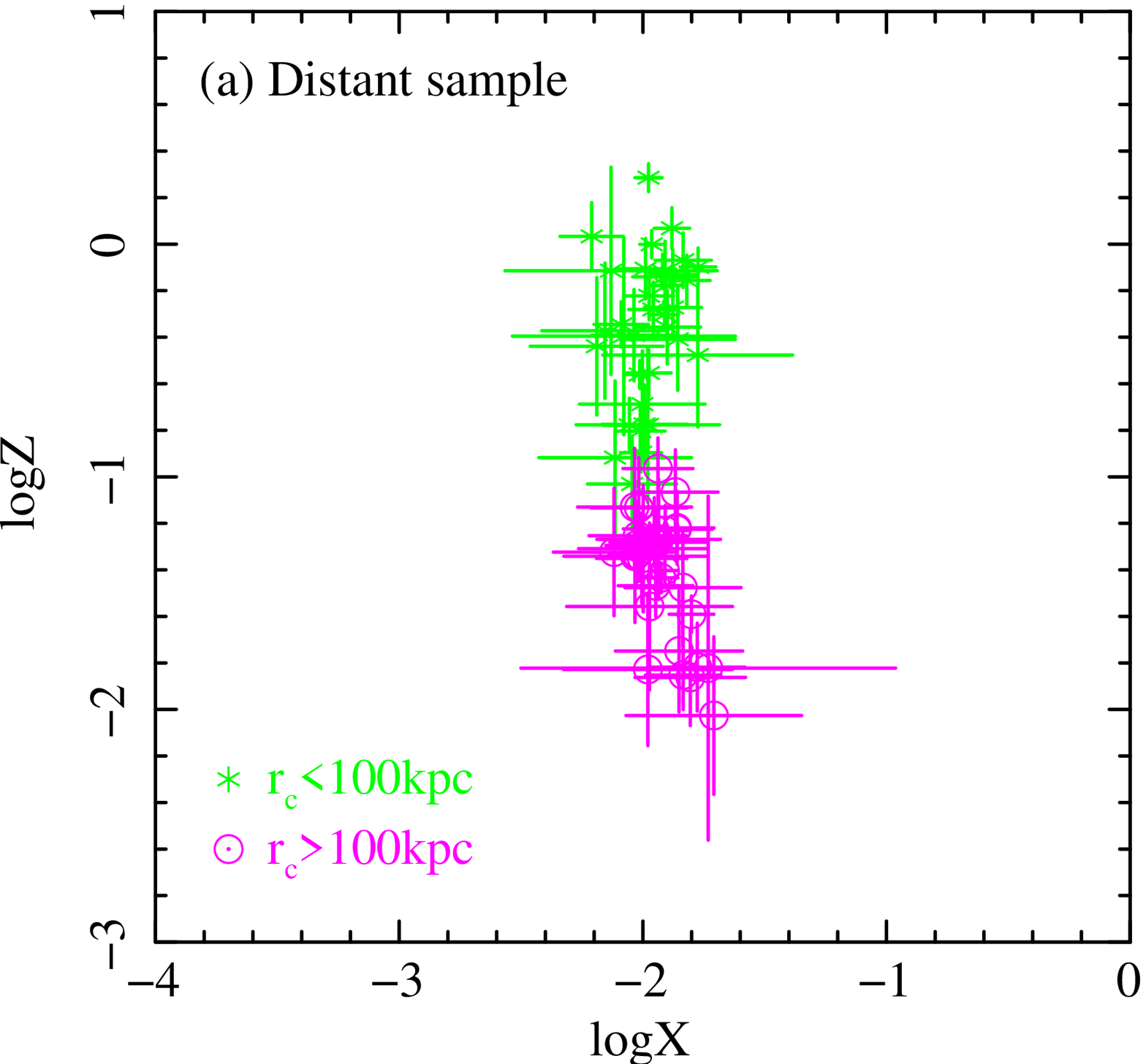}
        \includegraphics[scale=0.3]{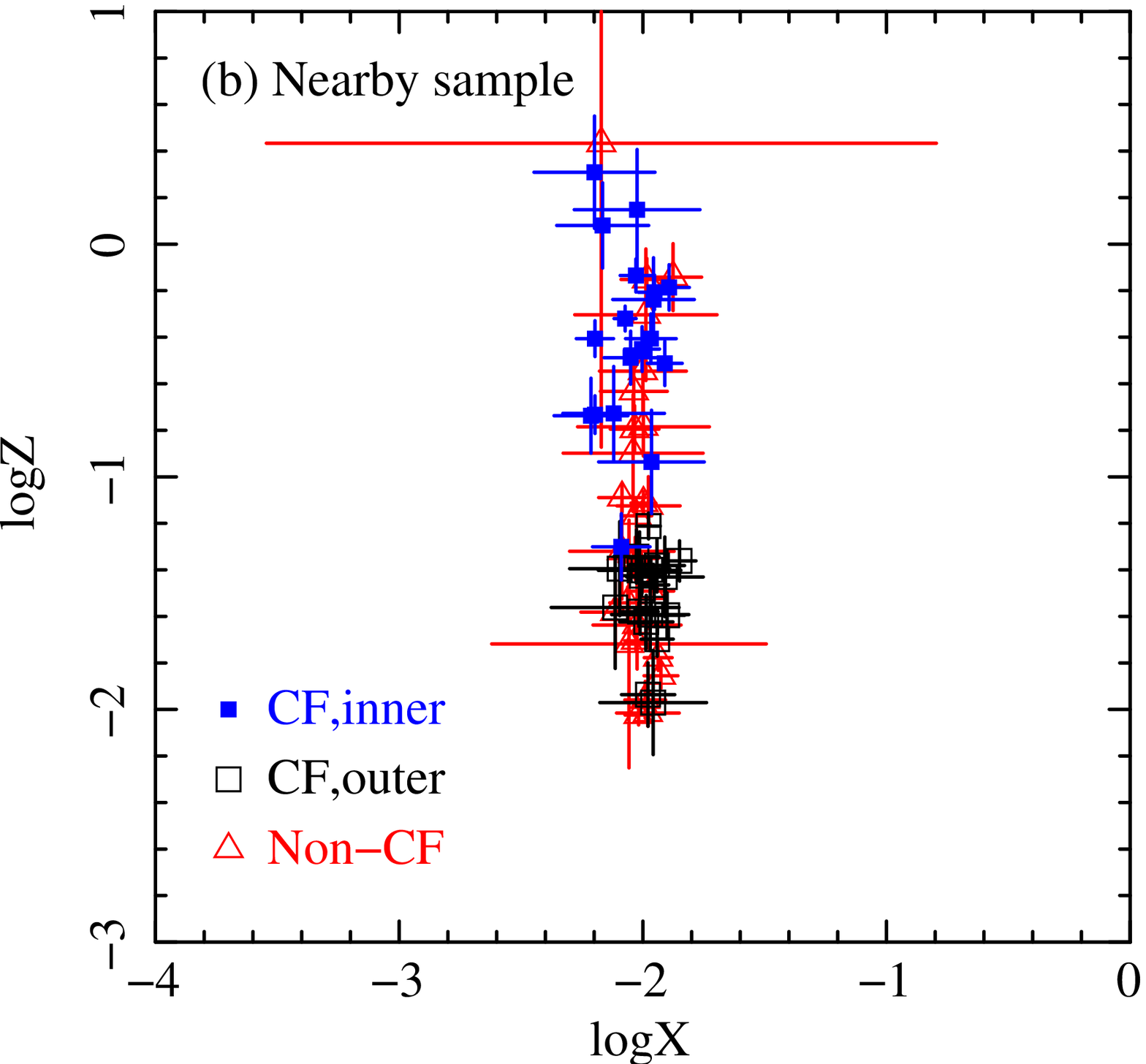}
        \vspace{-3mm}
    \caption{X-ray fundamental plane for the distant cluster sample \citep{ota06} and the nearby sample compiled by \citet{mohr99}. The distribution of the clusters projected onto the $\log{X}-\log{Z}$ plane is shown in each panel. For the nearby sample, according to Table~2 of \citet{mohr99}, non cooling-flow clusters are shown with the red triangles, and inner-core and outer-core components of cooling-flow clusters are separately shown with the solid blue boxes and open black boxes.}
        \label{fig:fp}
\end{figure}

An investigation of X-ray fundamental plane gives another clue to explore the
evolution of clusters. The presence of a planer distribution of nearby
clusters in 3-dimensional parameter space (the central gas density
$n_{r0}$, core radius $r_c$, and temperature $T$) was first noted
by \citet{fujita99}, implying that the clusters form a two-parameter
family. Applying this technique to distant clusters at $z>0.1$,
\cite{ota06} obtained the following three orthogonal parameters:
\begin{eqnarray}
X &\propto& n_{e0}^{0.44}r_c^{0.65} T^{-0.62} \\
Y &\propto& n_{e0}^{0.45}r_c^{0.44} T^{0.78} \\
Z &\propto& n_{e0}^{0.78}r_c^{-0.62} T^{-0.10} \label{eq:Zaxis}
\end{eqnarray}
and also confirmed the presence of the X-ray fundamental plane for the
distant cluster sample.  The distribution of clusters projected onto
the $X-Z$ plane is shown in Fig.~\ref{fig:fp}.  The $Z$-axis of the
plane is called the principal axis and represents the direction along
which the dispersion of the data points becomes the largest in the 3D
space. By setting $X\sim {\rm constant}$, Eq.\ref{eq:Zaxis} yields
$Z\propto r_c^{-1.78} T^{-0.10} \propto n_{e0}^{1.20}T^{-0.69}$. Since
the radiative cooling time is $t_{\rm cool}\propto
T^{1/2}n_{e0}^{-1}$, it is rewritten as
\begin{equation}
Z \propto t_{\rm cool}^{-1.2}.
\end{equation}
Therefore $t_{\rm cool}$ is considered to be a key parameter to
control the cluster's gas evolution. A trend of morphological change of
X-ray clusters along the $t_{\rm cool}$-axis is actually observed
\citep{ota06}. \citet{hudson10} noted that the cooling time is the
most suitable parameter to segregate CC/NCC clusters, which is in
agreement with the above result.

It should be noted that a cool core is also found in some irregular clusters. 
This phenomenon is interpreted as a remnant of a merging core and
may be used to diagnose the merging history \citep[e.g.,][]{markevitch07}.

\subsection{Beyond the $\beta$-model: observed gas profiles and the possibility of quasi-hydrostatic cooling}
To better reproduce observed X-ray surface brightness profiles, some
authors have introduced empirical models such as the double-$\beta$ model
\citep{jones84} and the modified $\beta$ model \citep{vikhlinin06}.
For the latter, they modified the original $\beta$-profile by adding a
cool density cusp at the center and steepening the slope at a large radius.
This model gave a good fit to spatially-resolved spectroscopic and
imaging data taken with {\it Chandra} for its large radial range
including the core emission \citep[see also][]{bulbul10}.

Given that radiative cooling plays an important role in the
thermodynamical evolution of ICM, how are the cool cores actually
formed and maintained?  The possibility of quasi-hydrostatic cooling
in the cluster core has been first noted by \citet{masai04}.  Unlike
isobaric cooling flows that increase the local density so the thermal
pressure $P(r)$ counteracts the local cooling, quasi-hydrostatic
cooling allows the gas to modify its profile or core size so $\nabla
P(r)$ balances the gravitational force.  The inflow is so moderate
that the hydrostatic balance is not disturbed significantly.  The
quasi-hydrostatic model predicts a temperature profile that approaches
a constant temperature of $\sim1/3$ that of ambient, non-cooling
gas, which agrees with those derived from X-ray observations of
relaxed clusters \citep{kaastra04,allen01,tamura01}.  Using a
hydrodynamics code, \citet{akahori06} investigated the evolution of
the core structure of radiatively cooling gas. They suggested a
radiative-cooling origin for the appearance of a small
($r_{c}\sim50$~kpc) core, while cooling is not important in clusters
with large cores. Their simulations also showed that the cluster core
maintains the quasi-hydrostatic condition before the initial central
cooling time has elapsed. This result gives a possible interpretation
on the observed double-peaked distribution of core size.

\citet{arnaud10} discussed the universal pressure profile for the REXCESS
cluster sample, and obtained the best-fit profile based on the
generalized NFW model by \citet{nagai07}. For the scaled temperature
and the density, \citet{arnaud10} found that their deviations from the
average scaled profile are anti-correlated with each other in the
core, $r/r_{500} < 0.2$ (figure 3 in their paper); the
anti-correlation is more clearly seen for cool core clusters. This
behavior is supported by the quasi-hydrostatic cooling picture.

Since the cooling time is shorter than the Hubble time for the CC
clusters, some heating is needed to sustain the system, otherwise it
would disappear $\sim$Gyr after becoming virialized. Practically,
however, heating due to mergers is likely invoked in the cluster's 
evolution. The clusters of core radii $>400$~kpc in the histogram
(Fig.~\ref{fig:rc_hist}) are attributed to mergers from their
irregular morphology. Recently, the process of cyclic evolution between CC and
NCC clusters was proposed by \cite{rossetti11} taking account of the 
lifetime of diffuse radio emission.

\subsection{Entropy profiles}
Measurement of a gas entropy profile provides important information on
the evolution of gas since it determines the structure of intracluster
gas and records the thermal history.
The gas entropy, $S$, in the field of cluster research is defined by 
\begin{equation}
	S = kT n_e^{-2/3}, 
\end{equation}
and is different from the original definition in the field of thermodynamics. 

The gravitational heating, namely conversion of the potential energy
to thermal energy, should depend on the depth of the gravitational
potential, which is approximated by the virial temperature of the
system. The entropy generation due to gravitational collapse is
predicted to be self-similar and follow a power-law form $S(r) \propto
r^{1.1}$ \citep{tozzi01,voit05}. Thus deviations from this {\it
  baseline} distribution may be attributed to cooling and heating
processes in the cluster. Earlier results on groups and clusters observed
with ROSAT showed that smaller systems like groups have entropy excess
called the ``entropy floor'' at the center while the slope of the
distribution follows the $\propto r^{1.1}$ law \citep{ponman03}. Thus
the non-gravitational effects, preheating or galaxy feedbacks, are
considered to play a greater role in smaller systems.

More systematic studies of entropy profiles with a large number of
clusters have been carried out; \citet[][]{cavagnolo09} derived radial
entropy profiles of ICM for 239 clusters with the {\it Chandra} data
(the ACCEPT sample) to find that most entropy profiles are well fitted
by a model consisting of a power-law plus a constant, $K(r) = K_0 +
K_{100}(r/100~{\rm kpc})^{\alpha}$.  The best-fit parameters are
$(K_0, K_{100},\alpha)=(16.1, 150, 1.20), (156, 107, 1.23)$ for
clusters with $K_0 \leq 50$ and $K_0 > 50~{\rm keV\,cm^2}$,
respectively.  They also showed that the distribution of central
entropy $K_0$ is bimodal, which peaks at $K_0\sim15~{\rm keV\,cm^2}$
and $\sim150~{\rm keV\,cm^2}$. A similar two peaked distribution has
been found in the REXCESS sample observed with {\it XMM-Newton}
\citep{pratt10}.  \citet{pratt06} measured the entropy profile in relaxed
clusters to find that outside $0.1r_{200}$ the scaled entropy profile
is consistent with gravitational heating while the scatter increases
with smaller radius and suggested that the results agree with models
of accretion shock.
  
\begin{figure}
    \centering
        \includegraphics[scale=0.28]{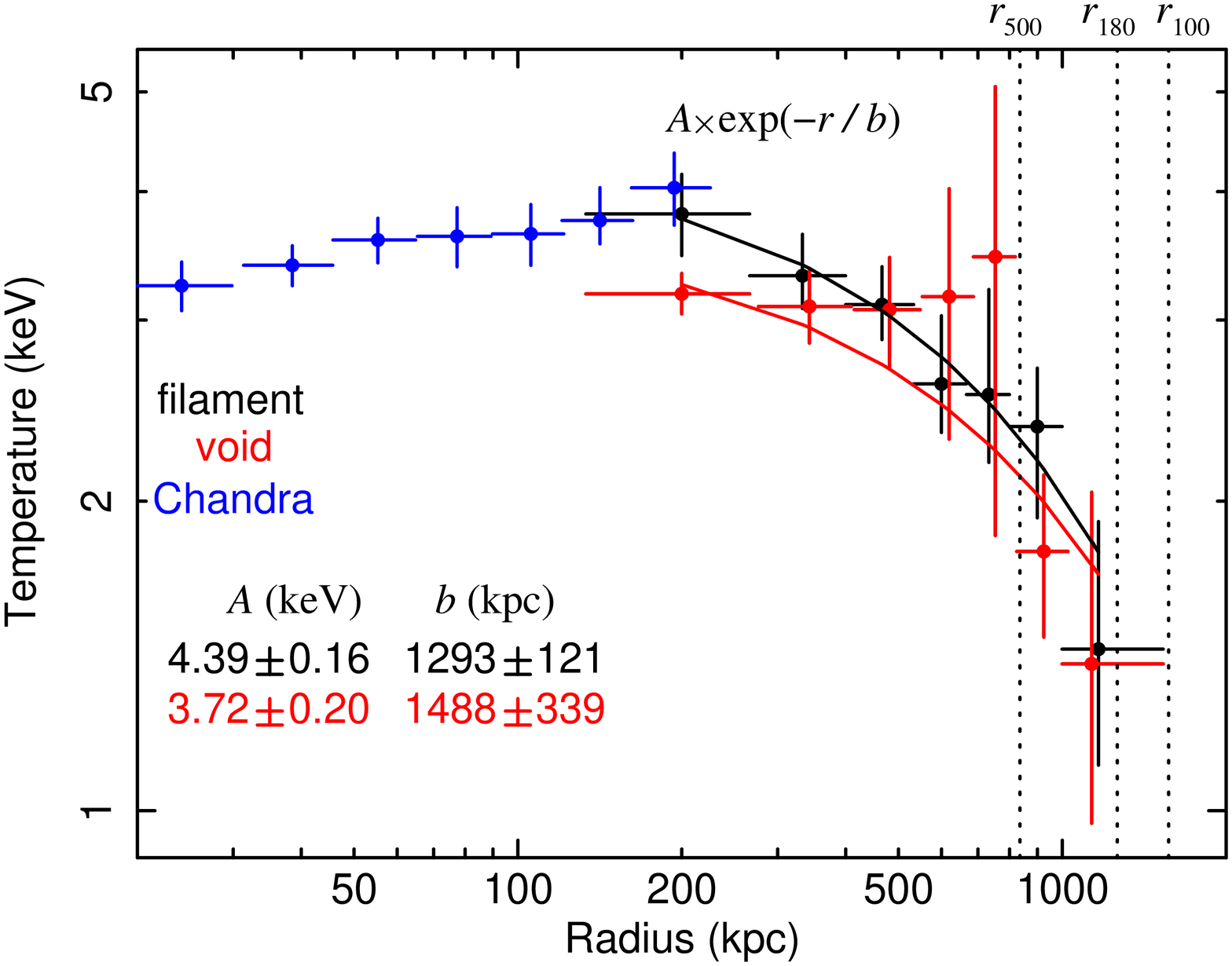}
        \includegraphics[scale=0.28]{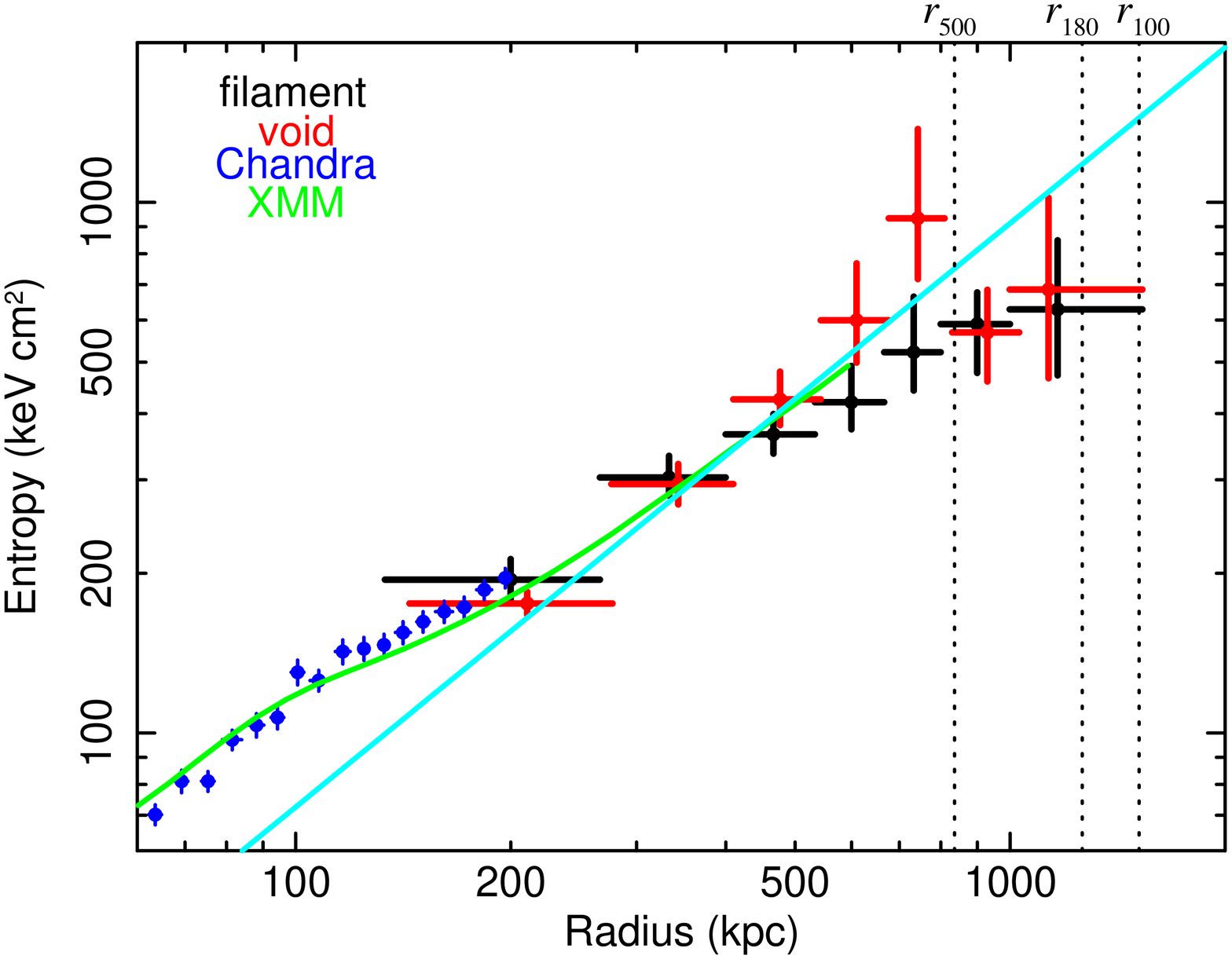}
    \caption{Radial temperature and entropy profiles of the Hydra A cluster measured with {\it Suzaku}\citep{sato12}. The results for filament (black) and void (red) directions are shown. The {\it Chandra} results \citep{david01} are plotted in blue. }
        \label{fig:hydra}
   \end{figure}

The advent of the {\it Suzaku} satellite \citep{mitsuda07} enables the
measurement of gas properties out to large radii because of its low
background level and high sensitivity. The temperature and entropy
distributions up to the virial radius have been derived for massive
clusters
\citep{geroge09,reiprich09,bautz09,kawaharada10b,hoshino10,simionescu11,akamatsu11}. The
latest result of Hydra A \citep{sato12} is shown in
Fig.~\ref{fig:hydra}. For those clusters observed with {\it Suzaku}, a systematic
drop in temperature by a factor of about three from outside the core to
$r_{200}$ was found and the entropy profiles become flatter beyond
$r_{500}$ in comparison with the $r^{1.1}$ profile. Some explanations
for observed low entropy are proposed and discussed: in-falling matter
retains some of its kinetic energy in the form of bulk motion \citep{kawaharad10b}, a gas
clumping effect \citep{simionescu11}, and deviation of electron
temperature from ion temperature \citep{akahori10}.

\section{Cluster merger and search for non-thermal phenomena}

According to the standard scenario of cosmic structure formation,
clusters are believed to have grown into their present shape via
collisions and mergers of smaller groups and clusters. A cluster
merger has a kinetic energy of the order of
\begin{equation}
E \sim \frac{1}{2}(M_1 + M_2 ) v^2 \sim 10^{65}~{\rm erg}\left(\frac{M_1+M_2}{10^{15}~M_{\odot}}\right)\left(\frac{v}{3000~{\rm km\,s^{-1}}}\right)^2,
\end{equation}
where $M_1$ and $M_2$ are masses of two objects. $v$ is the collision
speed and $v=3000~{\rm km\,s^{-1}}$ corresponds to the mach number of
2--3 in the intracluster medium. This is the most energetic event in
the Universe since the Big Bang. If two such objects collide with each
other under their mutual gravitational attraction, a huge amount of energy may
be released and a certain fraction is expected to heat the gas and
generate non-thermal particles through shock waves, and induce bulk
and turbulent gas motions.

We can recognize signatures of merging in many ways. In X-rays,
irregular morphology and the complex temperature structure of gas tell us
that the system is disturbed due to the past mergers. The most
prominent shock feature has been detected in 1E0657-56 (the Bullet
Cluster; \cite{markevitch02}). The bow shock propagates in front of a
bullet-like gas and significant jumps in temperature and density have
been found. The displacement between the peak positions of X-ray gas
and dark matter distributions have been identified in merging systems
such as the Bullet cluster and A2744 \citep{merten11}, which provide
an opportunity to constrain the self-interaction cross section of dark
matter particles.

\subsection{Gas bulk motion and turbulence}
In the course of merging, different portions of the hot gas are predicted to collide with each
other at a relative speed of $\sim$ a few $\times 1000~{\rm
  km\,s^{-1}}$, which will persist for several Gyrs after each merger
event \citep{norman99}. If the gas has a large bulk velocity compared
to its sound velocity, non-thermal pressure can no longer be neglected 
and has to be taken into account in estimating the cluster's 
mass. Suppose, for simplicity, that the gas is rigidly rotating with a
circular velocity of $\sigma_r (\propto r)$, then the balance against the
gravitational pull at radius $r$ on the rotational equatorial plane
then becomes,
\begin{equation}
-\frac{GM(r)}{r^2} = \frac{1}{\rho_{\rm gas}}\frac{\partial}{\partial r} P_{\rm gas} (1 + f\beta_r),
\end{equation}
where
\begin{equation}
\beta_r \equiv \frac{\mu m_p \sigma_r^2}{kT} \sim 1.07\left(\frac{\mu}{0.63}\right)\left(\frac{\sigma_r}{700~{\rm km\,s^{-1}}}\right)\left(\frac{kT}{3~{\rm keV}}\right)^{-1}
\end{equation}
and $f$ is the fraction of gas that is rotating
\citep{ota07}. Therefore the hydrostatic mass needs to be modified by a
factor of $(1+f\beta_r)$ given the presence of kinetic gas motion.

\begin{figure}
    \centering
        \includegraphics[scale=0.3]{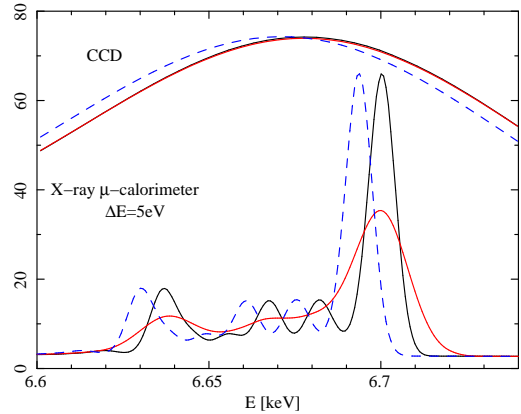}
        \caption{6.7~keV Fe XXV line profiles for $kT=5$~keV thermal
          gas convolved with the typical CCD and calorimeter detector
          responses. (i) thermal broadening only (black), (ii) thermal
          broadening + bulk motion $v_{\rm bulk} = 300~{\rm
            km\,s^{-1}}$ (blue), (iii) thermal broadening + turbulent
          broadening $v_{\rm turb} = 300~{\rm km\,s^{-1}}$ (red).}
        \label{fig:feline}
   \end{figure}

It is essential to constrain the gas motion through
observations. The cluster gas contains a large amount of heavy
elements such as iron, silicon, and oxygen etc.  If the gas has a
velocity along the line of sight, it produces Doppler shifts in
emission lines from the heavy ions. The line shift due to the
line-of-sight bulk velocity $v_{\rm turb}$ can be expressed as
follows
\begin{equation}
\Delta E_{\rm bulk} = E_0 \frac{v_{\rm bulk}}{c} = 6.7~{\rm eV}\left(\frac{E_0}{6.7~{\rm keV}}\right)\left(\frac{v_{\rm bulk}}{300~{\rm km\,s^{-1}}}\right), 
\end{equation}
where $E_0$ denotes the rest-frame energy of the line emission. For
example, a line shift due to the bulk velocity of $1000~{\rm
  km\,s^{-1}}$ corresponds to a shift in the 6.7~keV Fe-K line energy
by 22~eV. On the other hand, line broadenings due to turbulent and
thermal motions are given by the following two equations:
\begin{equation}
\Delta E_{\rm turb} = E_0 \frac{v_{\rm bulk}}{c} = 6.7~{\rm eV}\left(\frac{E_0}{6.7~{\rm keV}}\right)\left(\frac{v_{\rm turb}}{300~{\rm km\,s^{-1}}}\right)
\end{equation}
\begin{equation}
\Delta E_{\rm th} = E_0 \frac{\sqrt{kT/m}}{c} = 3~{\rm eV}\left(\frac{E_0}{6.7~{\rm keV}}\right)\left(\frac{kT}{5~{\rm keV}}\right)^{1/2}\left(\frac{m}{56m_p}\right)^{-1}
\end{equation}
Because the thermal width is inversely related to the ion mass $m$,
the contrast of turbulent broadening against the thermal one, $\Delta
E_{\rm turn}/\Delta E_{\rm th}$, becomes larger for larger $m$. Thus
the Fe emission line is the best-suited for velocity diagnostics in
clusters. Fig.~\ref{fig:feline} shows the Fe emission line model convolved
with instrumental responses assuming an X-ray CCD resolution of 130~eV
and an X-ray micro-calorimeter resolution of 5~eV (FWHM).

   By measuring the energy shift with X-ray spectroscopy, one can
   directly probe the dynamical state of the gas. However, it is not
   easy since it requires not only a high energy resolution and a good
   sensitivity but also a precise instrumental energy-gain
   calibration. Based on the careful assessment of positional gain
   variation of the {\it Suzaku} XIS detector \citep{koyama07}, a tight constraint on the
   bulk velocity with an accuracy of $700~{\rm km\,s^{-1}}$ has been
   placed in the central region of the Centaurus cluster
   \citep{ota07}. They placed the upper limit on the line-of-sight
   velocity difference as $1400~{\rm km\,s^{-1}}$. Hence, cluster mass
   estimation under a hydrostatic assumption is justified within a
   factor of about two--three. The Doppler-shift measurement using the Fe
   line has been carried out in several nearby clusters:
   \citet{sato08,sugawara09,sato11} derived the upper limit on the
   bulk velocity, and \citet{dupke07,dupke06} reported possible
   detection of bulk gas flow. Recently, the significant bulk velocity
   of a subcluster region relative to the main cluster, $\sim 1500~{\rm
     km s^{-1}}$, has been detected in A2256 by {\it Suzaku}
   \citep{tamura11}.

   The turbulent motion has been probed by measuring a
   spatially-resolved gas pressure map in the Coma cluster
   \citep{schuecker04}. The pressure fluctuation spectrum is found to
   be consistent with the Kolmogorov spectrum, yielding the lower
   limit of 10\% of the total gas pressure in turbulent
   form. The turbulent line broadening has been constrained using the
   Reflection Grating Spectrometer (RGS) on {\it XMM-Newton} in the central
   regions of ellipticals, groups and clusters \citep{sanders11}.
   They placed a strong upper limit on the turbulent motion
   ($<200~{\rm km\,s^{-1}}$) for several objects while line broadening
   has been found in Klemola 44 and a weak signature in
   RX~J1347--1145.

   Theoretical expectation for line shifting and broadening associated
   with turbulence and bulk motions as well as their detectability are
   discussed by \cite{sunyaev03,inogamov03,dolag05,pawl05}.

\subsection{Non-thermal hard X-ray emission}
At radio wavelengths, synchrotron emissions extending over a Mpc
scale have been discovered from more than 30 clusters \citep{giovannini99}. The existence
of radio halo emission suggests that relativistic electrons are being
accelerated in the intracluster space. Interestingly, there is a
correlation between the radio synchrotron power (non-thermal,
$P_{1.4}$) and X-ray luminosity (thermal, $L_X$) for merging clusters
while relaxed clusters without a radio halo lie in a region well
separated from the merging clusters on the $P_{1.4}-L_X$ plane
\citep{brunetti09}. It is suggested that generation of high-energy
particles is connected to the dynamical evolution of clusters
\citep{cassano10}.

In X-rays, the same population of high-energy electrons are thought to
interact with 3K CMB photons and then generate non-thermal
Inverse-Compton (IC) emission. The IC emission in excess of the
thermal emission is then predicted to be seen in the hard X-ray band
($>\sim10$~keV) where the thermal emission normally diminishes because
of the exponential cutoff in the continuum spectrum
(Eq.~\ref{eq:bremss}). In addition, from the radio observation alone,
we cannot separate the energy of magnetic fields from the energy of
high-energy electrons. However, by comparing the radio and hard X-ray
fluxes ($S_{\rm syn}$ and $S_{\rm IC}$), the cluster's magnetic field is
also estimated under the assumption that same population of
relativistic electrons scatter off of CMB photons since the ratio
$S_{\rm syn}/S_{\rm IC}$ is equal to the ratio between the energy
density of the magnetic field and the CMB
\begin{equation}
S_{\rm syn}/S_{\rm IC}=U_{B}/U_{\rm CMB}\label{eq:synchro-ic}
\end{equation}
\citep{rybicki86}. $U_B=(B^2/8\pi)$ and $U_{\rm CMB}=
4.2\times10^{-13}(1+z)^{4}~{\rm erg\,cm^{-3}}$. The exact derivations
of the synchrotron and IC emissions at a certain frequency are 
presented in \citet{blumenthal70}.

The existence of non-thermal IC hard X-rays in the Coma cluster has
been pointed out from {\it RXTE} \citep{rephaeli02} and {\it BeppoSAX}
observations \citep{fusco-femiano04}. Recent reports based on the
broad band X-ray observations with {\it Suzaku} \citep{wik09} and Swift
\citep{wik11} did not find any significant non-thermal hard X-ray
emission and the hard X-ray flux is reproduced by thermal models. This
mismatch among several satellites is suggested to be reconciled if
different sizes of field-of-views are taken into
consideration\citep{fusco-femiano11}.

\begin{figure}
    \centering
        \includegraphics[scale=0.3]{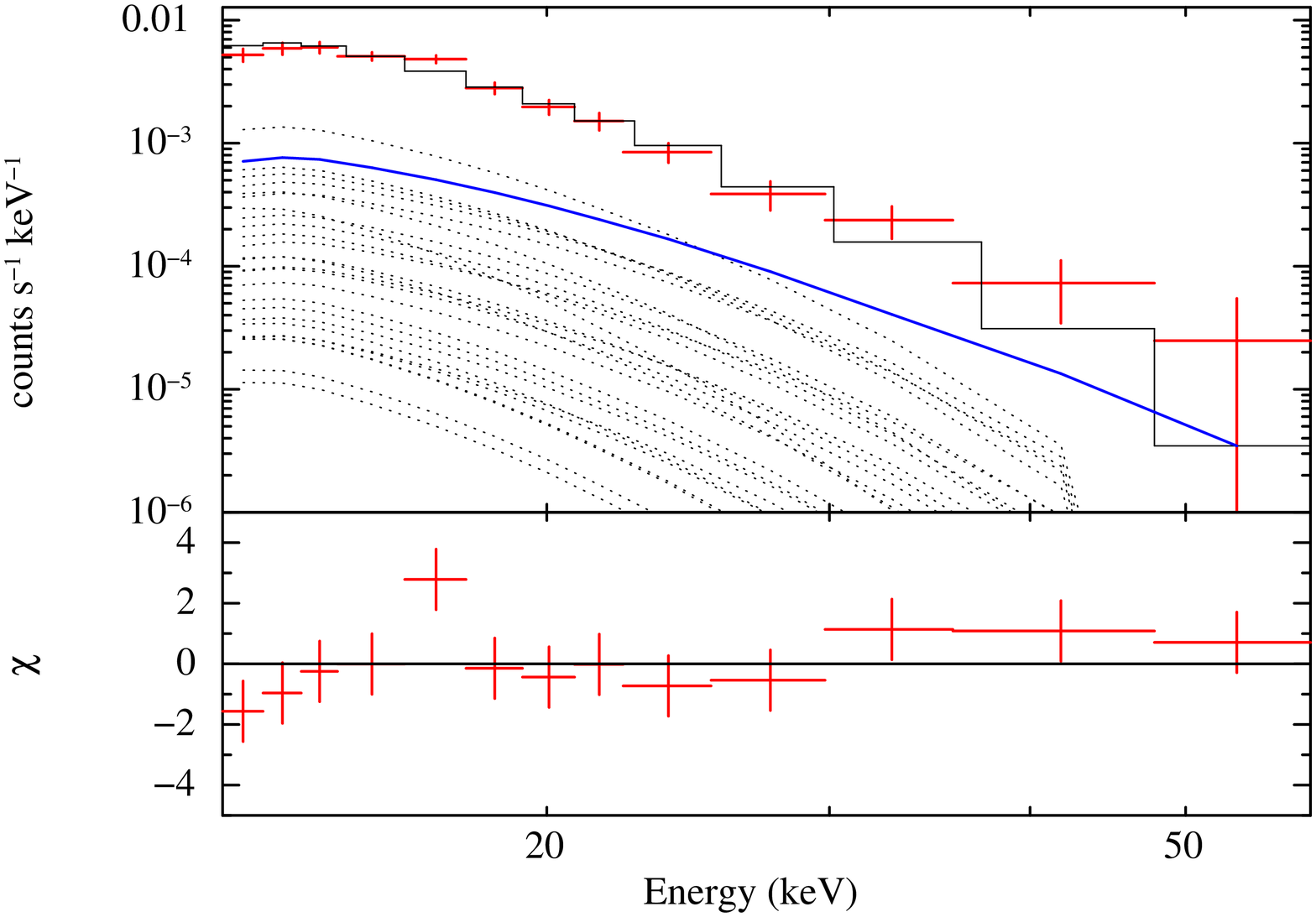}
        \caption{{\it Suzaku} HXD spectrum of the hot cluster of galaxies
          A2163. Significant emission is detected up to about
          $\sim 50$~keV. The spectral model consisting of
          multi-temperature thermal plasma (many thin dotted black
          lines) plus a non-thermal power-law component (blue) is
          indicated.}
        \label{fig:a2163}
   \end{figure}

\begin{figure}
    \centering
        \includegraphics[scale=0.3]{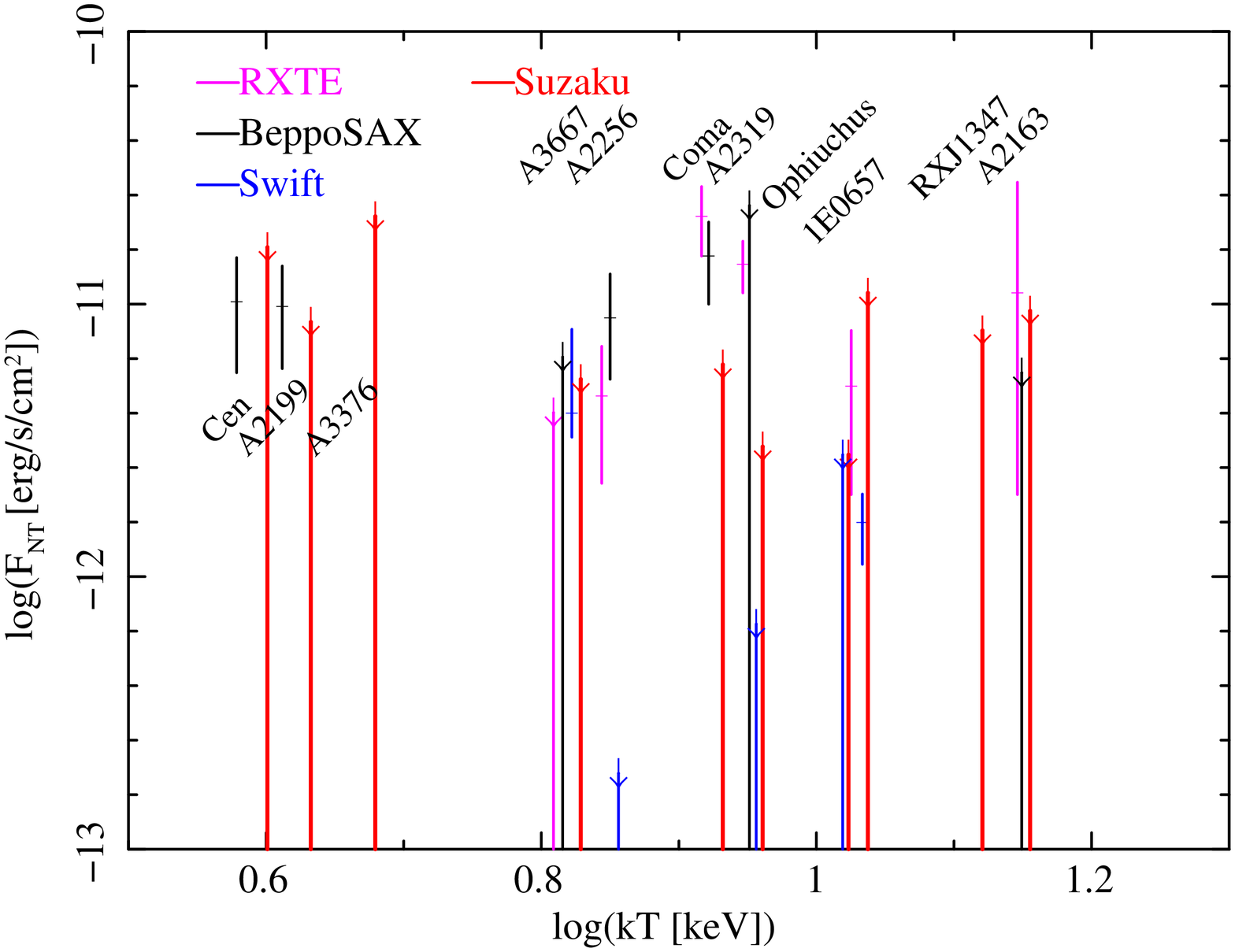}
        \includegraphics[scale=0.3]{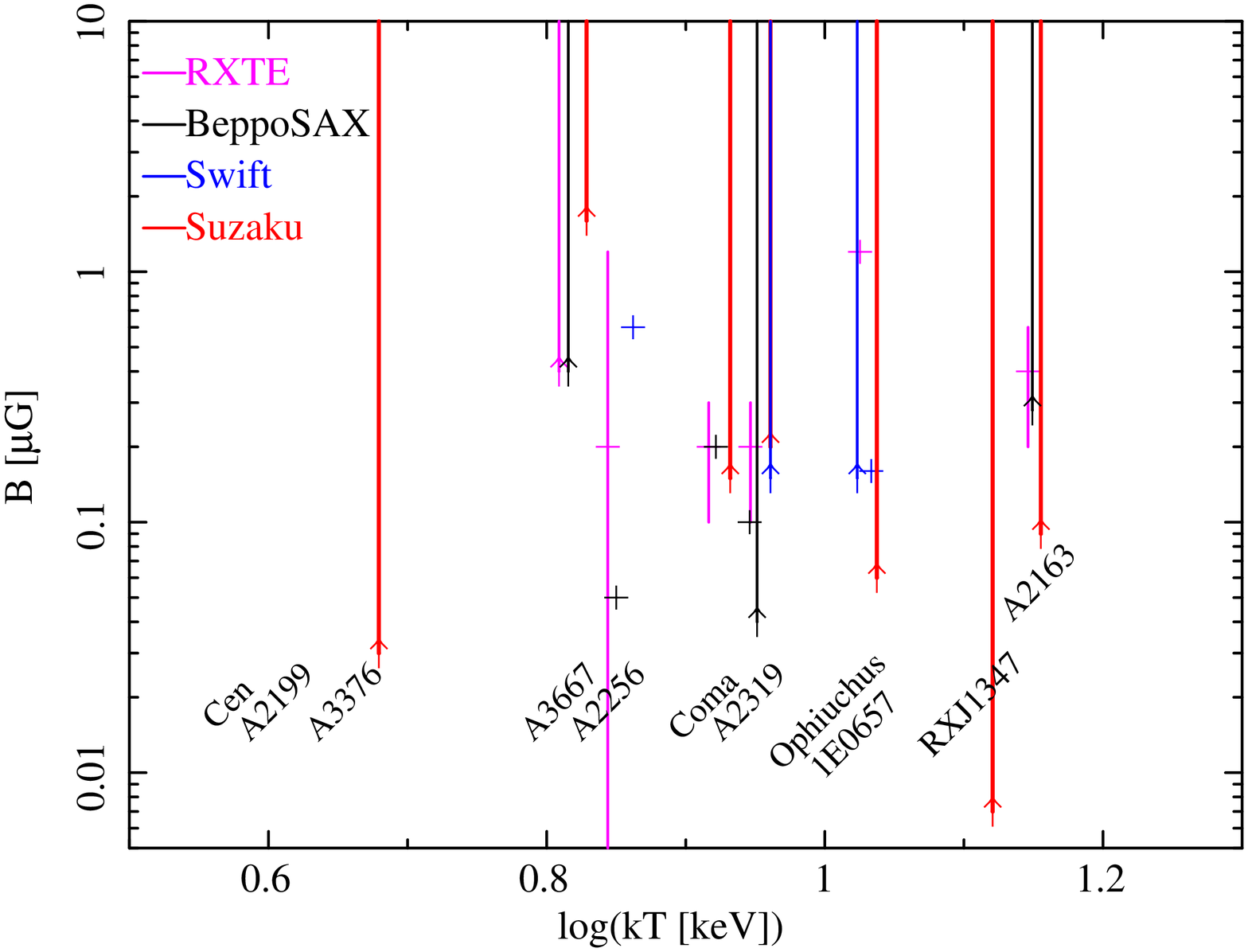}
        \caption{Non-thermal IC hard X-ray flux and cluster magnetic
          field in 10 clusters obtained with {\it Suzaku} (red). For {\it Suzaku},
          the results are quoted from
          \citet{kitaguchi07,fujita08,ota08,kawano09,nakazawa09,
            sugawara09, wik09, kawaharada10a}, Ota et al. in prep.,
          and Nagayoshi et al. in prep. For Swift (blue),
          \citet{ajello09,ajello10,wik11}. For {\it RXTE} (magenta) and
          {\it BeppoSAX}(black), see \citet{rephaeli08}; references
          therein.}
        \label{fig:nt}
   \end{figure}

Non-thermal hard X-ray emission has been constrained in about 10
bright clusters with {\it Suzaku}. The Hard X-ray Detector (HXD) on {\it Suzaku}
has a field of view of $34'\times34'$ (FWHM) at energies below 100~keV
and has achieved the lowest background level
\citep{takahashi07}. Fig.~\ref{fig:a2163} shows the hard X-ray
spectrum of the hottest Abell cluster A2163 ($z=0.203$) obtained with the 
{\it Suzaku} HXD. The additional power-law component does not significantly improve the
fit and the observed hard X-ray spectrum is well
explained by the multi-temperature thermal model, giving the upper
limit on the IC emission. This is consistent with the previous report
by {\it BeppoSAX} \cite{feretti01}.  The HXD results
\citep{kitaguchi07,fujita08,kawano09,nakazawa09, sugawara09,wik09,
  kawaharada10a} are compared with those from other satellites, {\it RXTE},
{\it BeppoSAX}, and Swift in Fig.~\ref{fig:nt}. There is no firm detection
of the IC emission reported for these 10 objects from {\it Suzaku}. The
cluster magnetic field obtained through the synchrotron-IC measurement
(Eq.~\ref{eq:synchro-ic}) based on the {\it Suzaku} HXD observations is also
plotted in the figure. Note that the estimation of magnetic field may
be affected by the assumption of index $p$ of the electron
distribution, $N(\gamma)=N_0\gamma^{-p}$ ($\gamma$ is the Lorentz
factor of the electron).

The situation of non-thermal X-rays from clusters remains uncertain,
and higher sensitivity in the high-energy range is required to further explore the physics of
gas heating and particle acceleration in clusters.

\subsection{Super-hot thermal gas in violent mergers}\label{subsec:super-hot}
To study thermal structure in clusters offers important perspectives
in understanding the merging configuration and heating process of the cluster's 
gas. Merger shock and evolution of temperature structure for ions and
electrons have been studied by numerical simulations. Given high sound
velocity in intracluster medium, it does not seem easy for an in-falling
sub cluster to acquire high Mach number to form strong shocks. On the
other hand, there are some pieces of observational evidence for
strongly heated gas that is likely to be generated by high-speed
($>2000~{\rm km\,s^{-1}}$) collisions.

\begin{figure}
    \centering
       \includegraphics[scale=0.35]{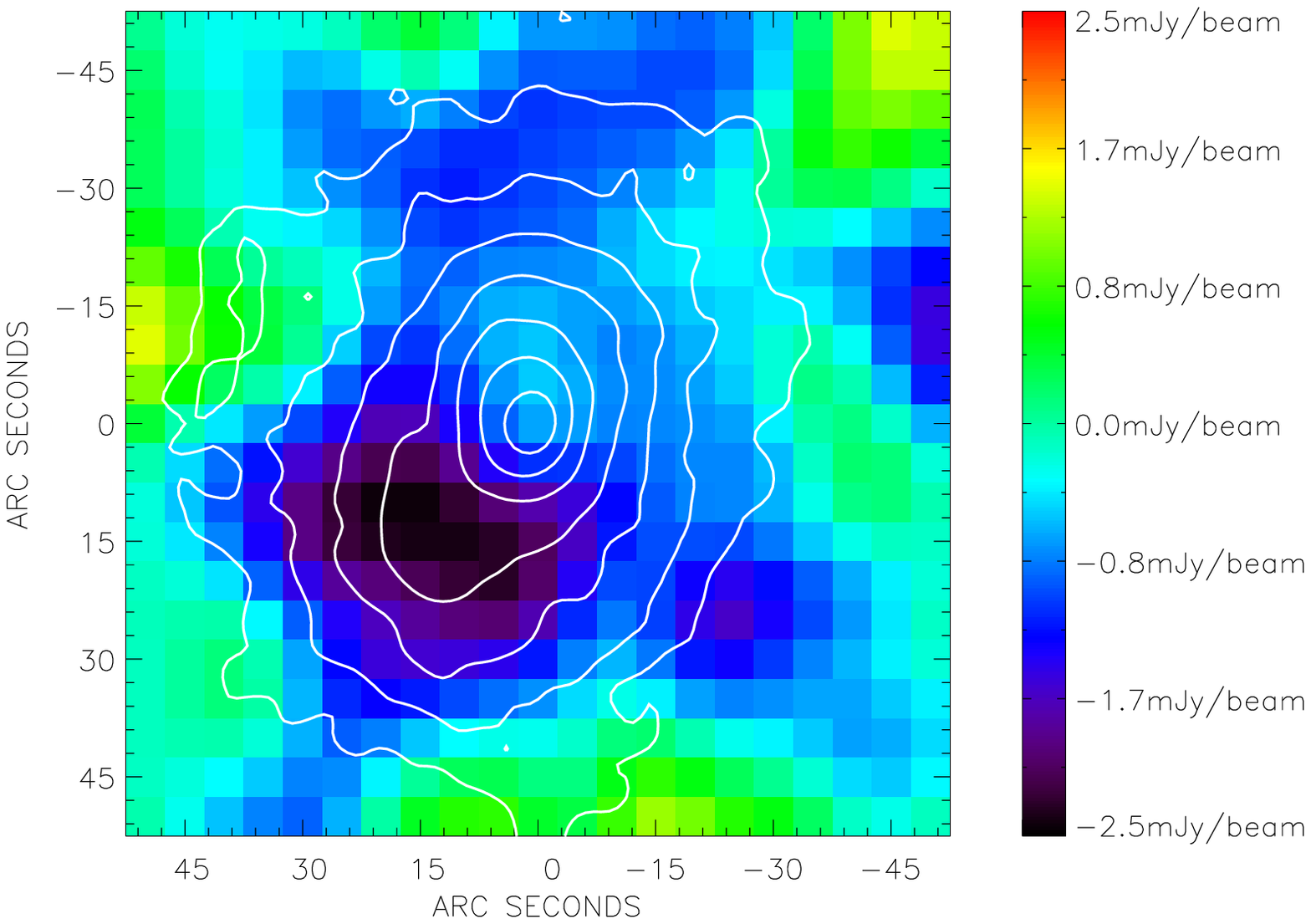}
       \includegraphics[scale=0.28]{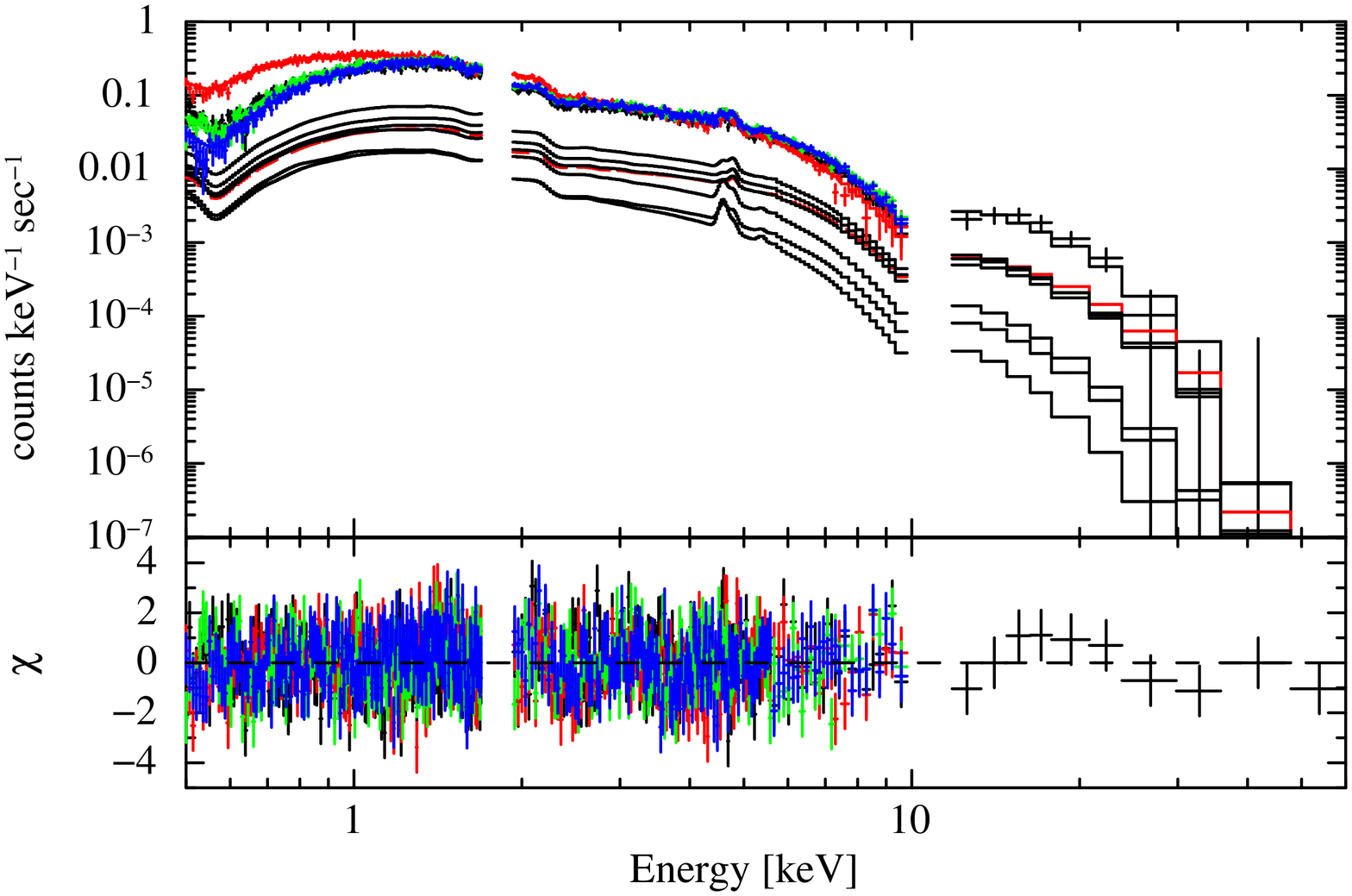}
       \caption{High-resolution SZ effect map taken at
         the 45-m Nobeyama telescope \citep[left;][]{kitayama04} and
         {\it Suzaku} broad-band spectra of the most X-ray luminous cluster
         RX~J1347--1145 \citep[right;][]{ota08}. The XIS data below
         10~keV and the HXD data above 10~keV are shown with 
         crosses. The step functions show the best-fit thermal model
         consisting of multi-temperature components for the ambient
         gas (many black lines) plus the very hot thermal gas (red
         line) identified in the South-East region of the cluster (see also Fig.~\ref{fig:rxj} right).}
        \label{fig:ota08}
   \end{figure}

The presence of extremely hot gas in the most X-ray luminous cluster
RX~J1347--1145 has been confirmed by the {\it Suzaku} broad band
spectroscopy \citep[Fig.~\ref{fig:ota08};][]{ota08}. From the joint
analysis of the {\it Suzaku} and {\it Chandra} data, the temperature of a hot clump
(Fig.~\ref{fig:rxj} right) is measured to be about $25$~keV, which is
more than two times higher than the surrounding gas. This unexpectedly
high-temperature gas has been pointed out previously by observations of the
Sunyaev-Zel'dovich(SZ) effect \citep{komatsu01,kitayama04},
and the broad-band X-ray data have improved the accuracy by
3-fold. Importantly, the X-ray spectrum of this hot component is more
accurately represented by a thermal emission model rather than a
non-thermal power-law model. The results support a scenario that this
cluster has experienced a recent violent merger as the very hot gas is
over-pressured and predicted to be short-lived ($\sim 0.5$~Gyr)
\citep{takizawa99}. It is also worth noting that the super-hot thermal
gas significantly contributes to the hard X-ray flux, which needs to
be precisely modeled in the search for non-thermal IC emission. Under
the detailed multi-temperature modeling of thermal emission
components, non-thermal IC emission is not found to be significant in
the hard X-ray spectra obtained with {\it Suzaku} for RX~J1347--1145
\citep{ota08}, Coma \citep{wik09}, Abell~2163 (Ota et al. in prep.) and the Bullet cluster
(Nagayoshi et al. in prep.).

   Is the very hot gas commonly seen in merging systems? In a nearby
   merging cluster A3667, a similar hot ($> 13$ keV) thermal component
   is suggested from the {\it Suzaku} observations
   \citep{nakazawa09}. Including a shock-front cluster, the Bullet
   cluster, the {\it Chandra} and {\it XMM-Newton} temperature maps show that some
   clusters contain very hot ($> 10$~keV) gas. The {\it Chandra} analysis by
   \citet{million09} indicated that the hard excess can be attributed
   to non-thermal gas or quasi-thermal gas with $kT > 20$ keV.

   The collision velocity necessary to explain such super-hot thermal
   gas due to strong shock heating is high ($\sim 3000-4000~{\rm
     km\,s^{-1}}$), which challenges the Lambda CDM model of cosmology \citep{lee10}.

\section{Future prospects}
X-ray spectroscopy and imaging observations bring us rich information
on the nature of galaxy clusters, not only the baryonic content but
also dark matter that governs the mass structure of the
objects. Large-scale cluster surveys in various wavelengths are now
on-going or planned, aiming to reveal the structural evolution in
the Universe and obtaining more stringent limits on cosmological
parameters. The baryonic mass fraction and cluster abundance as a
function of redshift have been used to constrain the dark matter and
dark energy densities as well as the dark energy equation of
state. These measurements require precise mass estimates of large
numbers of clusters, and thus understanding the physical state of
intracluster gas to calibrate scatter and redshift evolution and uncover any
bias in relationships between cluster mass and observables
\citep[e.g.,][]{majumdar03}.

Overall, clusters are {\it regular} objects, having positive
correlations between global quantities (gas temperature, bolometric
luminosity, gas mass etc) and the total mass derived either from X-ray
observations or a gravitational lensing effect. However, deviations from
the self-similar expectations have been observed in terms of the
power-law slopes and scatters around them. They are considered to have 
originated from non-gravitational effects like radiative cooling,
feedback from galaxies, bulk and turbulent gas motions, magnetic field
support etc.

Among these issues, measurement of velocity structure to high accuracy is
expected to be carried out by future high-resolution spectroscopy
using an X-ray micro-calorimeter.  The ASTRO-H satellite is scheduled to
be launched in 2014 \citep{takahashi10} and will play an critical role
in revealing the dynamics of clusters. The Soft X-ray Spectrometer
(SXS) onboard ASTRO-H is a non-dispersive spectrometer and enables
high-resolution (5~eV) observations for both point sources and diffuse
objects \citep{mitsuda10}. SXS will measure the kinetic gas motions to
an accuracy of $\sim 100~{\rm km\,s^{-1}}$ through observations of
line emissions. The Hard X-ray Imager on ASTRO-H \citep{kokubun10} will constrain the
non-thermal high-energy contents in clusters with its imaging
spectroscopy in the hard X-ray band. Now NuStar \citep{harrison10} is
successfully in orbit and draws peoples' attention to upcoming
observations with the first focusing telescope in the high energy
X-ray regime. The eROSITA on the Spectrum-Roentgen-Gamma mission will
perform an all-sky survey in the X-ray energy range and detect
$\sim100000$ clusters \citep{predehl10}. In conjunction with optical
and SZ surveys, the next-generation X-ray missions will largely
enhance the study of clusters and lead us to draw a more complete view of structure formation and evolution in the Universe.

\normalem
\begin{acknowledgements}
N.O. acknowledges the editors for giving me opportunity to write this review article.  
\end{acknowledgements}

\end{document}